\RequirePackage{fix-cm}
\documentclass{svjour3}                     
\smartqed  
\usepackage{graphicx}

\usepackage{multirow}
\usepackage{multicol}
\usepackage{booktabs}
\usepackage{algorithmicx,algpseudocode}
\usepackage[linesnumbered,ruled]{algorithm2e}
\usepackage{caption}

\usepackage{natbib}

\usepackage{xspace}
\newcommand{\OSS}{OSS\xspace}
\newcommand{\gh}{GitHub\xspace}
\newcommand{\discussions}{Discussions\xspace}
\newcommand{\GHD}{\gh \discussions \xspace}

\begin{document}

\title{Looking for related discussions on \textit{GitHub Discussions}}



\author{\textit{Márcia Lima}*         \and
        Igor Steinmacher \and
        Denae Ford \and
        Evangeline Liu \and
        Grace Vorreuter \and
        Tayana Conte \and
        Bruno Gadelha
}

\authorrunning{Lima et al.}


\institute{*\textit{Corresponding Author}: Márcia Lima \at
              Federal University of Amazonas (UFAM) \& Amazonas State University (UEA)\\
              Tel.: +55 92 3305-1181\\
             \emph{Present address: Av. Rodrigo Otavio - 6200. Manaus-AM-Brazil} \\  
             \email{marcia.lima@icomp.ufam.edu.br;msllima@uea.edu.br} 
             \and
             Igor Steinmacher \at
            Federal Technological University of Paraná (UTFPR) \\
            \email{igorfs@utfpr.edu.br}
            \and
            Denae Ford \at
            Microsoft Research\\
            \email{denae@microsoft.com}
            \and
            Evangeline Liu \& Grace Vorreuter \at
            GitHub\\
            \email{evi-liu;gracevor@github.com}
            \and
            Tayana Conte \& Bruno Gadelha \at
            Federal University of Amazonas (UFAM)\\
            \email{tayana;bruno@icomp.ufam.edu.br}
}




\date{Received: date / Accepted: date}

\maketitle

\begin{abstract}
Software teams are increasingly adopting different tools and communication channels to aid the software collaborative development model and coordinate tasks. Among such resources, Programming Community-based Question Answering (PCQA) forums have become widely used by developers. Such environments enable developers to get and share technical information. Interested in supporting the development and management of Open Source Software (OSS) projects, GitHub announced \emph{GitHub Discussions} --- a native forum to facilitate collaborative discussions between users and members of communities hosted on the platform. As GitHub Discussions resembles PCQA forums, it faces challenges similar to those faced by such environments, which include the occurrence of related discussions (duplicates or near-duplicated posts). While duplicate posts have the same content---and may be exact copies---near-duplicates share similar topics and information. Both can introduce noise to the platform and compromise project knowledge sharing. In this paper, we address the problem of detecting related posts in GitHub Discussions. To do so, we propose an approach based on a Sentence-BERT pre-trained model: the {\tt RD-Detector}. We evaluated {\tt RD-Detector} using data from different OSS communities. OSS maintainers and Software Engineering (SE) researchers manually evaluated the {\tt RD-Detector} results, which achieved 75\% to 100\% in terms of precision. In addition, maintainers pointed out practical applications of the approach, such as merging the discussions' threads and making discussions as comments on one another. OSS maintainers can benefit from {\tt RD-Detector} to address the labor-intensive task of manually detecting related discussions and answering the same question multiple times.




\end{abstract}

\section{Introduction}
\label{sec_introduction}


Software engineering teams actively adopt social resources to support the collaborative software development and coordinate team members' tasks~\citep{storey2014r,storey2016social,tantisuwankul2019topological}. Teams use such resources to communicate, learn, answer questions, obtain and give feedback, show results, manage, and coordinate activities~\citep{storey2016social}. E-mail, chats, and forums are examples of collaborative social media that support software teams' communications~\citep{storey2014r,storey2016social,perez2018collaborative}. In recent years, Programming Community-based Question Answering (PCQA) has increasingly attracted different users' attention and has become widely used by software developers~\citep{wang2020duplicate,pei2021attention, ford2018somentoring}. Developers rely on PCQA to quickly find answers for technical questions~\citep{mamykina2011design,zhang2017detecting}, impacting the software development process~\citep{yazdaninia2021characterization}.

Concerned with helping the development and management of OSS projects hosted on GitHub, in 2020, the company announced \GHD \citep{Evi20,hata2022github}. In order to be ``A new way for software communities to collaborate outside the codebase''~\citep{Niyogi20}, the \GHD forum provides opportunities for OSS communities to interact and discuss project-related issues collaboratively. The \discussions forum is a place where OSS communities can talk about work, ask questions, plan new releases, request code reviews, make announcements, disclose information, recruit contributors, get insights into the project, feature important information, or simply chat~\citep{hata2022github, Evi20}.

However, the PCQA type forums, like any other question and answering forums, should be concerned about the quality decay of their contents. \cite{silva2018duplicate} point out that repeated questions put at risk that quality. Prior research handles this problem by focusing on automating the duplicate question detection in PCQA forums~\citep{zhang2015multi_DupPredictor,ahasanuzzaman2016mining,zhang2017detecting,silva2018duplicate,wang2020duplicate,pei2021attention}. Research that aims to detect duplicate questions in Stack Overflow highlights that duplicates can (1) pollute the platform with already-answered questions~\citep{silva2018duplicate}; (2) consume the time of experts, as they must manually analyze and look for duplicates~\citep{zhang2017detecting,wang2020duplicate}; and (3) make users wait unnecessarily for answers to questions that had been already asked and answered~\citep{ahasanuzzaman2016mining}.

Although both \GHD and Stack Overflow resemble a PCQA forum, they differ regarding the communicative intentions behind the posts. Posts on \GHD cover a software project ecosystem~\citep{hata2022github} (on GitHub, each project has its own forum). Stack Overflow questions cover a broader context; they aim to answer developers' technical questions regardless of a specific software project. However, text evidence collected from \GHD threads (discussion first posts and comments) shows that the forum, like Stack Overflow, contains duplicate and related posts\footnote{https://github.com/homebrew/discussions/discussions/1531}\footnote{https://github.com/homebrew/discussions/discussions/707}\footnote{https://github.com/vercel/next.js/discussions/22211}. Currently, maintainers manually manage duplicates on the \GHD  forum. Based on their previous knowledge, maintainers identify duplicates as new discussions arise. However, manual strategies are less effective and are subject to human subjectivity and imprecision~\citep{ahasanuzzaman2016mining,zhang2017detecting,wang2020duplicate}.

Different approaches have already been proposed and evaluated to solve the problem of duplicates in PCQA forums. Such methods often use pre-labeled datasets to train or optimize the duplicates detection process. However, the posts on \GHD are project-related, and they are not previously categorized using pre-defined topics. In addition, different project contexts emerge every time an OSS community starts to use the Discussions forum. Therefore, creating a manual labeled sample is costly and can always be considered inefficient due to human imprecision, and small, given the diversity of projects hosted on GitHub. Thence, we did not train or optimize domain-specific machine learning models. The dynamic growth of the \GHD and the diversity of projects hosted on the platform create opportunities and challenges to develop an approach to detect related discussions on GitHub Discussions.

Therefore, motivated by the context mentioned, this work aims to propose the {\tt RD-Detector}, an automated approach to detect related discussions in the \GHD forum. More specifically, related discussions are those duplicated or near-duplicated ones. Duplicate posts have the same content and could be exact copies. Meanwhile, near-duplicates have similar topics and share similar information. In this work, we consider duplicates a subtype of related discussions for practical purposes. From now on, we use the term 'related discussions' to refer to duplicate and near-duplicate discussion posts.
More specifically, our research question (RQ) is:

\textbf{RQ:} Are general-purpose deep machine learning models to Natural Language Processing (NLP) problems effective in detecting related discussions in OSS communities? 

To do so, we propose the {\tt RD-Detector} approach.  {\tt RD-Detector} uses a Sentence-BERT (SBERT) pre-trained general-purpose model to create sentence embedding representations of discussion posts and compute the semantic similarity between pairs of discussions. The approach outputs a set of related discussion candidates. The RD-Detector differs from the previous approaches as (1) it does not rely on pre-labeled datasets; (2) it does not rely on specific models to understand OSS communities' discussions context; (3) it can assess different software contexts hosted on GitHub; and (4) it aims for precision in detecting related discussion posts. OSS maintainers and Software Engineering (SE) researchers evaluated the {\tt RD-Detector} results. They both classified the pairs of related discussion candidates as related or not and pointed to evidence on the challenges and practical applications in detecting related discussion posts in the GitHub Discussions.

Our results show that we can use a general-purpose machine learning model for detecting related discussions in OSS communities. We assessed {\tt RD-Detector} over different datasets, achieving 75\% to 100\% in terms of precision. We identified the imprecision of the term `related discussions' as a limitation during the evaluation. The general-purpose machine learning model brought flexibility to the approach. As an advantage, we can point out that the results does not restrict to a specific software project context. OSS maintainers can benefit from the results to minimize the work overhead in manually detecting related threads and the rework in answering the same question multiple times. Besides, the results support OSS maintainers to tackle the platform pollution and to address the project knowledge sharing degradation, as it occurs in different discussions threads.

The main contributions of this research include:
\begin{itemize}
    
\item The {\tt RD-Detector}, an approach based on deep machine learning models to detect related posts in GitHub Discussions.

\item Empirical evidence on using a general-purpose machine learning model to detect related discussions held by OSS communities.

\item Empirical evidence regarding the {\tt RD-Detector} practical applications in OSS communities, from OSS maintainers' perspective.

\end{itemize}

\section{Background}
\label{sec_background}

OSS projects are highly distributed environments usually composed of self-directed development teams~\citep{chen2013knowledge,tantisuwankul2019topological}. Therefore, knowledge sharing is a critical factor for the success of OSS teams \citep{chen2013knowledge,tantisuwankul2019topological}. \cite{chen2013knowledge} point out that `knowledge sharing is an interactive cuing process in which knowledge provided by one team member becomes the cue for other members to retrieve relevant but different knowledge stored in their own memory.' Given that the \GHD is a collaborative communication channel where OSS communities make questions, debate, and announce project-related issues, we conjecture that the Discussions is an online environment that promotes knowledge sharing in OSS communities. \cite{tantisuwankul2019topological} point out that the key to software projects hosted on collaborative platforms such as GitHub success are the communities' interaction and the project knowledge sharing.

This section presents the main concepts used in this research and previous works on detecting duplicates in PCQA forums. Section~\ref{sec_backgroung:GH_Discussion} describes the \GHD forum (our object of study). Section~\ref{sec_background:related_work} presents the concept of duplicates from the researchers' perspective and how they tackle duplicates or related posts in programming Q\&A forums.

\subsection{\textit{GitHub Discussions}}
\label{sec_backgroung:GH_Discussion}

\GHD is a feature to any public or private repository on GitHub \citep{Evi20}. It facilitates collaborative discussions among maintainers and the community for a project on GitHub \citep{Evi20}. GitHub company suggests using the Discussions forum to ask and answer questions, share information, make announcements, and lead or participate in project-related conversations \citep{discussion}. The Discussions is a collaborative communication forum for OSS maintainers, code and non-code contributors, newcomers, and users to discuss projects' use, development, and updates in a single place without third-party tools~\citep{resources_GH,discussion}. In addition, the Discussions stands out for being a place to distinguish day-to-day conversations and conversations aimed at engineering teams (Issues or Pull Request)~\citep{hata2022github}.

Users, maintainers, contributors, and newcomers can join in a conversation by creating, commenting, reacting, or reading a discussion post~\citep{discussion}. The Discussions users can also search selected topics in discussion posts~\citep{discussion_search}. To do so, they must specify keywords in the GitHub search engine. Users can restrict the search results to the discussion title, body text, or comments by applying correct qualifiers.

\begin{figure}[h]

  \centering
  \includegraphics[width=\linewidth]{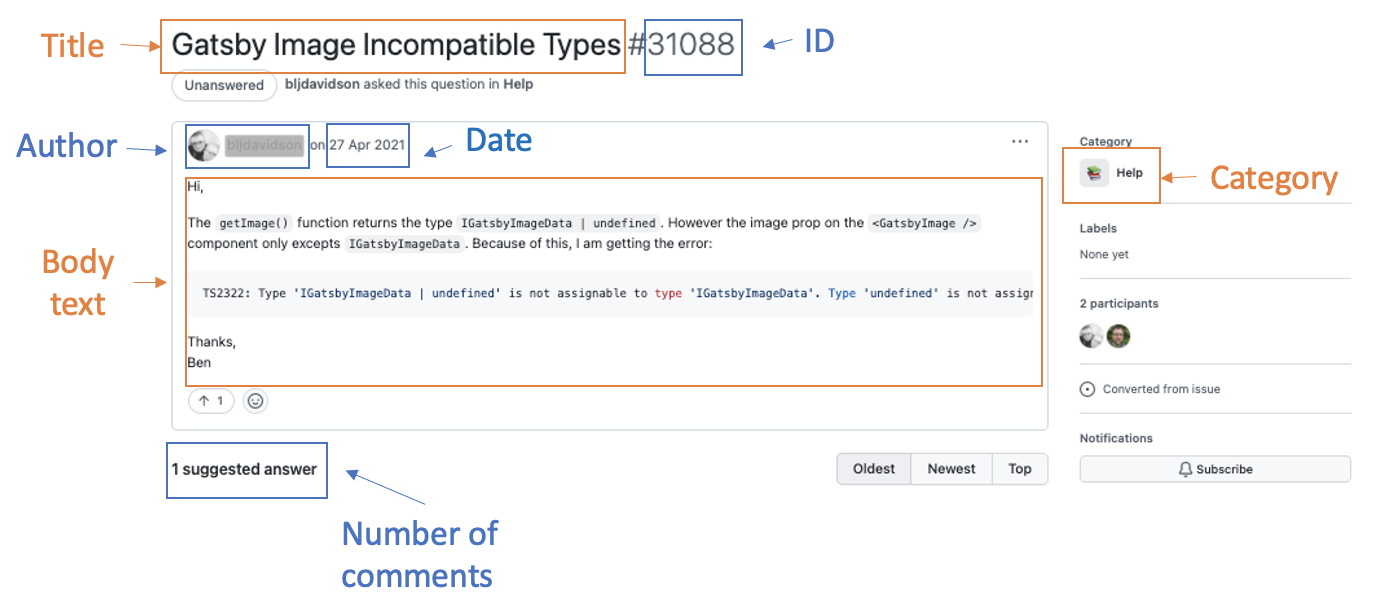}
  \caption{An example of a discussion post.}
   \label{fig:oneDiscussion}
\end{figure}

Authors must specify the discussion title, body text, and category to create a new discussion (Figure~\ref{fig:oneDiscussion}). The category is a mandatory attribute of a discussion post. It helps organize conversations into predefined classes, allowing community members to chat in the right place and find discussions with similar characteristics~\citep{discussion_category}. Authorized members can define, create, or delete categories in a repository according to the project needs. By default, the \GHD forum provides five types of categories: {\tt Announcements}, {\tt Q\&A}, {\tt Ideas}, {\tt Show and tell}, and {\tt General} \citep{discussion_category}. Maintainers can create Announcements discussions to share project updates and news. Users can create {\tt Q\&A} discussions to ask questions, suggest answers, and vote on the most appropriate feedback. {\tt Ideas} discussions can report or share ideas regarding the project improvements. {\tt Show and tell} discussions discuss relevant creations, experiments, or tests. Finally, {\tt General} discussions address any issue relevant to the project~\citep{discussion_category}.

Maintainers report a positive acceptance of \GHD in the OSS communities~\citep{Evi21_mainteiners}. They highlight that the forum enables the growth of the communities in the same place they use to collaborate, improve, and increase the OSS community members' engagement, and separate the issues trackers from questions, feature requests, and general chatting~\citep{Evi21_mainteiners}. In addition, maintainers point out they can use the discussions' threads to access historical data~\citep{Evi21_mainteiners}. They can keep track of the questions already asked, their proposed solutions, and the suggestions made. With this threading support, they can individually address demands without losing them in broader discussions.

\cite{Evi20} highlights the use of \GHD by five selected \OSS communities: {\tt Dogecoin},\footnote{https://github.com/dogecoin/dogecoin/discussions} {\tt NASA},\footnote{https://github.com/nasa/cFS/discussions} {\tt Next.js},\footnote{https://github.com/vercel/next.js/discussions} {\tt Pixar},\footnote{https://github.com/PixarAnimationStudios/OpenTimelineIO/discussions} and {\tt React}.\footnote{https://github.com/reactwg/react-18/discussions} The {\tt Dogecoin} community uses \GHD  to centralize its developers' conversations in one place, {\tt NASA} uses it to collaborate with the OSS community on its core flight systems. The {\tt Next.js} community uses Discussions to plan features, promote community interaction, and exchange ideas about the project. The {\tt Pixar} project uses the forum to engage the OSS community responsible for creating its {\tt OpenTimelineIO} API. Finally, {\tt React} started using discussions to introduce a new release and give collaborators a place to ask questions about it.

This work uses public discussion posts collected from three OSS communities to evaluate our approach. Our dataset comprises the discussion posts of {\tt Gatsby}, {\tt Homebrew}, and {\tt Next.js} communities (Table~\ref{tab:GH_repositories}).

\subsection{\textit{Related Work}}
\label{sec_background:related_work}

Several studies aim to detect duplicates on PCQA forums by proposing different approaches for duplicate question detection \citep{zhang2015multi_DupPredictor,ahasanuzzaman2016mining,zhang2017detecting,silva2018duplicate,wang2020duplicate,pei2021attention}. Generally, researchers model the duplicate detection task as a ranking problem, a binary classification problem, or a combination of both. Ranking problems combine features to sort the top-k most similar documents (discussions, texts, questions) \citep{liu2011learning} and predict duplicates. Classification problems classify pairs of documents into predefined categories (e.g., duplicates or non-duplicates) \citep{ahasanuzzaman2016mining,zhang2017detecting}. 

The literature characterize duplicate posts on PCQA forums as:
(1) questions that address the same topic, but are not necessarily identical copies \citep{silva2018duplicate,wang2020duplicate};
(2) questions conceptually equivalent to other questions previously posted \citep{abric2019can}; (3) questions that were already asked and answered before \citep{zhang2017detecting,wang2020duplicate}; (4) questions asked to solve the same problem \citep{ahasanuzzaman2016mining}; and (5) questions that `express the same point' \citep{zhang2015multi_DupPredictor}. We can note the duplicates conceptualization is not a rigid definition; it can depend on human subjective criteria. \cite{ahasanuzzaman2016mining} point out that exact duplicates are significantly rarer, and many users duplicate questions by asking the same thing in different ways.

\cite{zhang2015multi_DupPredictor} proposed {\tt DupPredictor}, an approach to detect potential duplicate questions in Stack Overflow. The {\tt DupPredictor} combines the similarity scores of different features. The authors evaluated the approach's effectiveness using a pre-labeled Stack Overflow dataset, which achieved 63.8\% of recall rate. \cite{ahasanuzzaman2016mining} and \cite{zhang2017detecting} addressed duplicate detection as a supervised classification problem. The authors used a pre-labeled dataset to train and validate a classifier to detect duplicates in Stack Overflow and measured the proposed approachs' effectiveness using the recall rate. \cite{silva2018duplicate} implemented the {\tt DupPredictorRep} and {\tt DupeRep} approaches. The authors adopted an official Stack Overflow dataset dump to train, optimize, and evaluate the approaches that predict and classify questions as duplicate or not, respectively. \cite{wang2020duplicate} used deep learning techniques to detect duplicate questions in Stack Overflow. The authors proposed three different approaches based on Convolutional Neural Networks (CNN), Recurrent Neural Networks (RNN), and Long Short-Term Memory (LSTM) to identify duplicates. The authors also used a pre-labeled Stack Overflow dataset to train, evaluate, and test the classifiers and measured the approaches' effectiveness using the recall rate. Finally, \cite{pei2021attention} proposed an Attention-based Sentence pair Interaction Model (ASIM) to predict the relationship between Stack Overflow questions. The authors used a Stack Overflow dump to train the software engineering specif domain. The proposed model achieved 82.10\% precision and 82.28\% recall rates, outperforming the baseline used.

Conversely, \cite{abric2019can} analyzed how helpful duplicate questions are to the software development community. The authors argue that duplicates on Stack Overflow help the developer community by providing different formulations of the same problem or solution. Also, the additional answers may be more understandable for some users.

Previous work has also highlighted the problem of duplicates in issue reports and pull requests (PRs) on GitHub \citep{li2017detecting,yu2018dataset,li2018issue,ren2019identifying,li2020redundancy,zhang2020ilinker}. \cite{li2018issue} analyzed explicit links in both issues reports and PRs. They reported the importance of such links in identifying duplicates. The authors analyzed 70,686 links that represented duplication relationships, from which 59.03\% identified duplicate issues and 40.97\% identified duplicate pull requests. \cite{li2020redundancy} presented empirical evidence on the impact of duplicate PRs on the software development effort. They observed that the 'inappropriateness of OSS contributors' work patterns and the drawbacks of their collaboration environment would' result in duplicates. In addition, researchers have been proposing different approaches to address duplicates in GitHub \citep{li2017detecting,yu2018dataset,ren2019identifying,zhang2020ilinker}. \cite{li2017detecting} and \cite{ren2019identifying} proposed automatic approaches based on traditional Information Retrieval (IR) and NLP techniques to detect duplicates in pull requests. \cite{yu2018dataset} constructed a dataset of historical duplicate PRs extracted from projects on GitHub - the {\tt DupPR} dataset - by using a semi-automatic approach. \cite{zhang2020ilinker} proposed the {\tt iLinker}, an approach to detect related issues in GitHub. The authors trained {\tt iLinker} to learn the embedding corpus and models from the project issue text.

The mentioned approaches to detect duplicates on PCQA forums have two common points: the use pre-labeled datasets to learn or better understand the domain application, and their effectiveness is measured using the Recall rate. However, this research faces some challenges that make it different from the previously mentioned papers: (1) there is not a pre-labeled dataset of duplicated or related discussions, extracted from \GHD, through which we can train or optimize machine learning models; (2) creating a pre-labeled base of related discussions is a time-consuming task and can always be considered inefficient and small; and (3) there is a vast diversity of repositories hosted on GitHub, which provide different software project contexts (in January 2020 the platform already hosted more than 200 million repositories\footnote{https://github.com}). Different from the mentioned strategies to detect duplicates in issue reports and pull requests, our approach is based on general-purpose deep learning models to address the topic duplication problem in Discussions. Therefore, in order to detect related discussion candidates (duplicate or near-duplicate) in the \GHD forum, our proposed approach (1) does not rely on a pre-labeled dataset; (2) does not rely on domain-specific models; (3) applies to different OSS project domains; (4) is based on a Sentence-BERT pretrained model, and (4) aims at improving precision rates rather than the recall rates. In Section~\ref{sec_approach}, we propose a method to detect related discussions in GitHub Discussions.

\section{ {\tt RD-Detector}:  Related Discussions Detector}
\label{sec_approach}

In this section, we present the {\tt RD-Detector} approach. The {\tt RD-Detector} aims to detect pairs of related discussion candidates in collaborative discussion forums of OSS communities. To this end, the {\tt RD-Detector} scores the Semantic Textual Similarity (STS) between discussion posts to find discussions with similar meanings. Using the STS value, we can detect exact copies (duplicates) and discussions that share the same topic (near-duplicates). The greater the similarity values, the greater the chances of duplicates. We consider any duplicated or near-duplicated discussions as `related discussions.' The {\tt RD-Detector}'s output is a set of related discussion candidates, which, from now on, we denote $R$.

\begin{figure}[h]

  \centering
  \includegraphics[width=\linewidth]{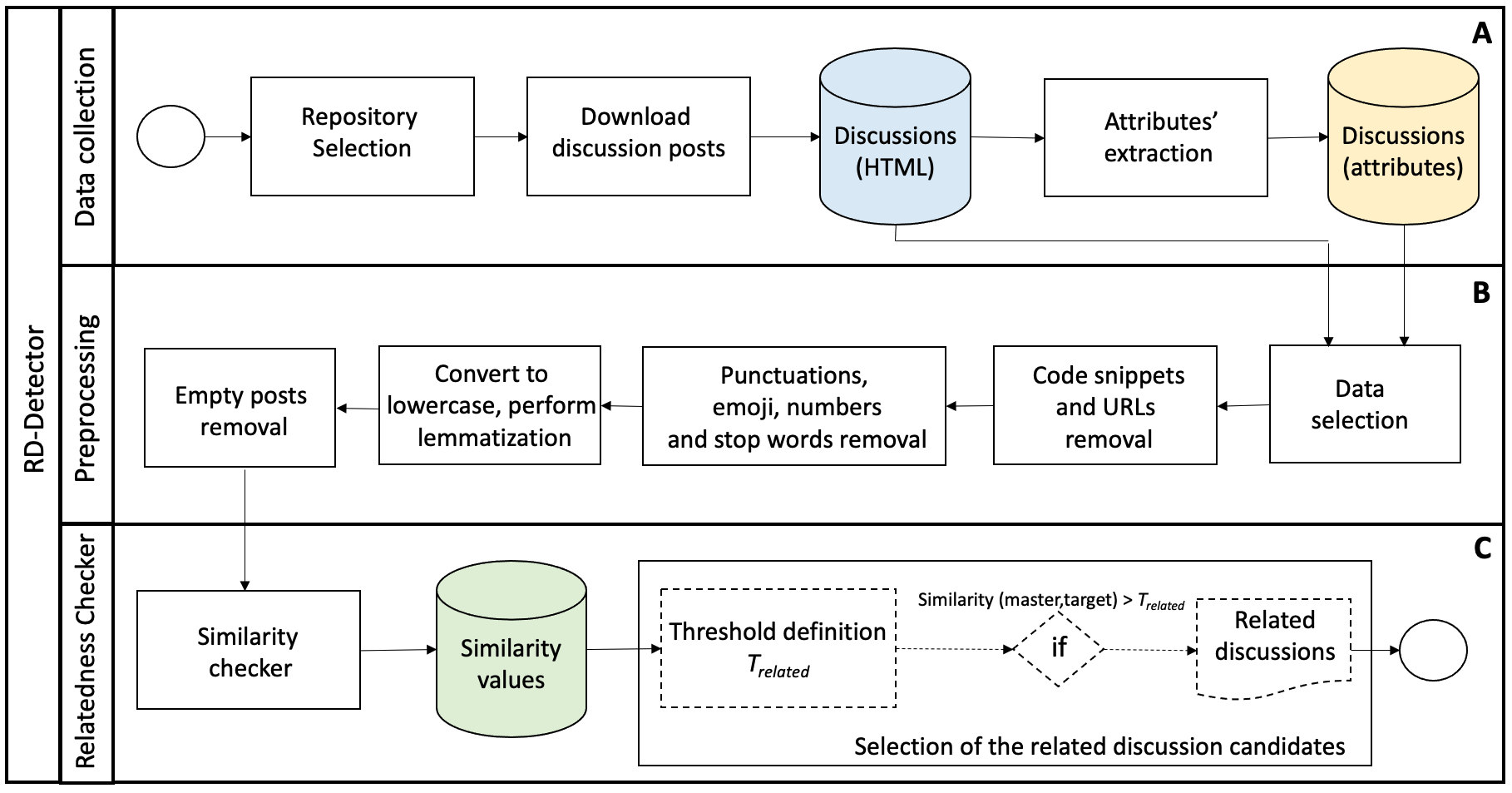}
  \caption{The {\tt RD-Detector} approach}

   \label{fig:overall_process}
\end{figure}

Figure~\ref{fig:overall_process} shows the overall process conducted to detect the sets of related discussion candidates. The process comprises three phases:  Data collection (Figure~\ref{fig:overall_process} - A), Preprocessing (Figure~\ref{fig:overall_process} - B), and Relatedness Checker (Figure~\ref{fig:overall_process} - C). We describe the phases of the proposed approach as follows.

\subsection{Data collection}

In the ``Data collection'' phase, we focus on gathering data from OSS communities. {\tt RD-Detector} was assessed using data collected from public discussions threads held in three \GHD forums (Table~\ref{tab:GH_repositories}).

Our data collection comprises three steps (Figure~\ref{fig:overall_process} - A). In the first step, we select the OSS repositories of interest. Next, we collect the discussions threads of the selected repositories. Finally, we extract some preselected attributes from each discussion thread. The {\tt RD-Detector} considers the title and the body of the discussion posts to calculate the semantic text similarity values between pairs of discussion posts. We do not consider comments because they are feedback or design reasoning about the discussions' main topics. During the data collection phase, the {\tt RD-Detector} extracts the author, date, and category type of the discussion posts. The {\tt RD-Detector} uses the attributes to select the discussion posts and create sub-datasets according to predefined filters (Section~\ref{sec_caseGH:prepocessing}).

\subsection{Preprocessing}
\label{sec_approach:preprocessing}

Applying text similarity algorithms in discussion content requires data preprocessing to optimize the algorithm execution~\citep{zhou2017machine}. We executed a preprocessing pipeline to remove noisy information and format the dataset collection (Figure~\ref{fig:overall_process} - B). The preprocessing phase comprises five steps. In the ``Data selection'' step, the {\tt RD-Detector} selects the title and the body text of the discussion' posts. In addition, in ``Data selection'' we can filter discussion posts according to the values of their attributes. We can select discussion posts that fulfill certain conditions, such as category type (using the category attribute).

The following two preprocessing steps, ``Code snippets and URL removal'' and ``Punctuation, emoji, numbers, and stopwords removal,'' aim to remove noise data from the content of the discussions. We remove the code snippets since the machine learning model used is a general-purpose model trained to handle NLP problems. Furthermore, the structure and vocabulary of natural language texts differ from machine codes~\citep{sirres2018augmenting}. We use HTML tags, e.g. {\tt <code>}, to remove code snippets and URLs.

The ``Punctuation, emoji, numbers, and stopwords removal'' step removes punctuations, emojis, numbers, and stopwords from discussions content. To this end, we use the Python Natural Language Tool Kit (NLTK) library~\footnote{https://www.nltk.org/}. We do not remove the following punctuation symbols `.' and `\_' because they concatenate words, e.g., `Next.js' and `version\_1.3'. In the next preprocessing step, we convert the remaining text to lowercase and apply lemmatization~\citep{zhang2017detecting}. Finally, we eliminate zero-length discussion posts in the ``Empty posts removal'' step.

\subsection{Relatedness Checker}

The ``Relatedness Checker'' phase is the core of the {\tt RD-Detector} (Figure~\ref{fig:overall_process} - C). In this phase, the approach computes the similarity score of discussions' pairs and detects the related discussion candidates. The preprocessed data is the input of this phase (Algorithm~\ref{algo:match}, lines 0-3). The set R, containing pairs of related discussion candidates, is the output of the ``Relatedness Checker'' phase (Algorithm~\ref{algo:match}, line 30). This phase comprises two steps: (1) the similarity measurement (Similarity checker) and (2) the Selection of related discussion candidates. The steps are as follows.

\subsubsection{Similarity checker}
\label{sec_approach:similarityChecker}

We use a semantic text similarity (STS) checker~\citep{agirre-etal-2015-semeval} to score the similarity between pairs of discussion posts. The similarity checker measures the similarity score using a general-purpose Sentence-BERT pre-trained model \citep{reimers2019sentence}.

The {\tt RD-Detector} uses the public {\tt all-mpnet-base-v2}\footnote{https://huggingface.co/sentence-transformers/all-mpnet-base-v2} model to compute semantically significance sentence embeddings of each discussion post. The model maps sentences and paragraphs to a 768-dimension dense vector space. At the time of this work execution, the {\tt all-mpnet-base-v2} model provided the best quality in sentences embeddings computing and semantic searching performance~\citep{SBERT_site}.

We compare the sentences embeddings using the cosine-similarity score. We used the score to set the similarity value between pairs of discussions and to rank sentences with similar meaning ($similarity(master, target)$). The cosine similarity value outcome is bounded in [0,1]~\citep{abric2019can}. Cosine similarity value 1 refers to identical discussion contents, value 0 refers to dissimilar discussions~\citep{abric2019can}. There is no hierarchy or priority relationship between pairs of masters and targets discussions. Masters discussions are the oldest post, and targets ones are the most recent. Therefore, the similarity value for ($master_i$,$target_j$) is equal to the similarity value for ($target_j$,$master_i$). 

Researchers train and optimize custom-built machine learning models using labeled dataset samples. We propose an approach based on a general-purpose machine learning model to address the lack of a pre-labeled dataset and generalize the outcomes to different OSS communities. One can benefit from general-purpose models to detect related discussions on the GitHub Discussions forum because of the lack of an official dump of related posts to train or optimize automatic approaches and the significant number of OSS projects ecosystems hosted on the platform. General-purpose models application is not restricted to a single context, as they are trained and optimized in different datasets. Specifically, the {\tt all-smpnet-base-v2}'s base model is the microsoft/mpnet-base model. Experts optimized the {\tt all-smpnet-base-v2} model over 1 billion sentence pairs collected from various datasets (Reddit comments (2015-2018), WikiAnswers Duplicate question pairs, Stack Exchange Duplicate questions (titles+bodies), etc.)~\citep{hugginface}. Context-specific models are trained and optimized according to pre-labeled samples that summarize context specifics, not guaranteeing effectiveness when used outside the training context. In addition, general-purpose models usage provide advantages such as (1) no need for local complex computational structures for training, validation, and testing of models, (2) no need for model parametrization, and (3) no need for model retraining, revalidation, and retest whenever the context change \citep{polyzotis2017data,zhou2017machine,schelter2018challenges,lee2020machine}. The dynamism with which OSS communities grow justifies general-purpose machine learning model usage. Indeed, public machine learning models bring flexibility to the {\tt RD-Detector} approach. As experts release new exchangeable machine learning models, one can update the {\tt RD-Detector}.

Therefore, for each pair of a $(master, target)$ discussion, the {\tt RD-Detector} computes the semantic similarity value that captures the relatedness of the posts, Algorithm~\ref{algo:match} - lines 4-13. The similarity values are the input to the ``Selection of related discussion candidates'' step.

\begin{algorithm}[h]
\SetKw{Let}{let}

 \KwIn{Set of discussion posts \textbf{$D$}}
 \KwIn{Set of discussion attributes \textbf{$\Delta$}}
 \KwIn{\textbf{K} value}
 \KwIn{Sentence-BERT \textbf{model}}

\ForEach{discussion $d_{i} \in $ D'} {
       $D'\gets D'+ preprocessing(d_{i}) $\;
  
}  

\texttt{/* Similarity Checker */}\\
\ForEach{ discussion $d_{i} \in D'$} {
\ForEach{discussion $d_{j}, (d_{j} \neq d_{i}) \in D'$} {
       $master_{i} \gets d_{i}$\;
       $target_{j} \gets d_{j}$\;
       $similarity\_value_{i\_j} \gets similarity(master_{i},target_{j}) $ \textit{/* Using \textbf{model} */} \;
       $tuple \gets (master_{i},target_{j},similarity\_value_{i\_j}) $\;
       $save (tuple,similarity\_values\_file) $\;
  
}  
}
\texttt{/* Threshold definition */}\\
$S \gets \{\}$ \;
{\ForEach{discussion $d_{i} \in D'$} {
    $master_{i} \gets d_{i}$\;
    $S = S + topK\_sim\_values(master_{i},K) $\;
    }
}
$Q1 \gets 25th\_percentile(S)$\;
$Q2 \gets 50th\_percentile(S)$\;
$Q3 \gets 75th\_percentile(S)$\;
$IQR \gets Q3 - Q1$\;
$T_{related} = Q3 + (1.5 * IQR)$\;
\texttt{/* Selection of related discussion candidates */}\\
$R \gets \{\}$ \;
\ForEach{$ master_{i},target_{j},similarity\_value_{i\_j} \in similarity\_values\_file $}
{

        
    \lIf {$ (similarity\_value_{i\_j} \geq T_{related})$}
       {$R \gets R + (master_i,target_j)$}

}
{\Return $R$ \;}
\caption{Identifying Related Discussions}
\label{algo:match}
\end{algorithm}

\subsubsection{Selection of related discussion candidates}
\label{sec_approach:selection}

We rank the related discussion candidates according to their similarity values. We do not use a pre-determined threshold, $x$ to filter the related discussions. Instead, the {\tt RD-Detector} computes what we call `local threshold,' denoted by $T_{related}$ based on each project dataset. The $T_{related}$ is defined using descriptive statistics that describe the characteristics of a specific similarity value set. The local threshold use allows the adaptation of the proposed approach to different OSS communities' contexts. We set the $T_{related}$ value following four steps:

\begin{enumerate}

    \item \textbf{Defining the $K$ value:} The $K$ value delimits the search bounds for related discussion candidates. The {\tt RD-Detector} uses the $K$ value to select the similarity values of the top-K most similar discussions to each post in the input dataset.The greater the value of $K$, the greater is the number of similarity values selected and the search bounds. Setting $K=5$, the approach uses the similarity values of the top-5 most similar discussions to every post in the dataset. Setting $K=10$, the {\tt RD-Detector} selects the similarity values of top-10 most similar discussion posts. One can specify different $K$ values. $K$ is an input value (Algorithm~\ref{algo:match}).

    \item \textbf{Creating the distribution $S$:} The $S$ distribution is a collection of similarity values.  $S$ contains the similarity values of the $K$ most similar target discussions for each discussion post in the dataset  (Algorithm~\ref{algo:match}, lines 14- 19).

    Let $n$ be the number of discussion posts in the dataset, $K$ the number of the most similar target discussions to every discussion in the dataset, and $value\_ij$ the similarity value of a given $master_i$ and $target_j$ discussion pair, the distribution $S$ is:

    \begin{center}

    $S = <value_{1\_1}, value_{1\_2},...,value_{1\_K}, value_{2\_1},$ \\ $value_{2\_2}, ...,value_{2\_K},...,value_{n\_1}, value_{n\_2},...,value_{n\_K}> $ \\
    
    \end{center}

    \item \textbf{Determining the descriptive statistics of $S$:} We use descriptive statistics variability measures to understand how dispersed the distribution is. To this end, we calculate the interquartile range ($IQR$), along with the 25th percentile ($Q1$), the 50th percentile ($Q2$), and the 75th percentile ($Q3$), Algorithm~\ref{algo:match} - lines 20 to 23. Next, we find the Upper Inner Fence value (Equation~\ref{eq:upper_inner_fence}) that identifies the outliers in $S$~\citep{tukey1977exploratory}.
    
     \begin{equation}
        \label{eq:upper_inner_fence}
         \textit{Upper inner fence} =  Q3 + (1.5 * IQR)
    \end{equation}

    \item \textbf{Setting the local threshold ($T_{related}$):} Because we assume that the greater the semantic similarity value of a pair of discussion posts, the greater the chances they are related discussion, we set the local threshold to the upper inner fence value (Algorithm~\ref{algo:match} - line 24). Therefore, we have that:
    
       \begin{equation}
        \label{eq:limiar}
         T_{related} = \textit{Upper inner fence}
    \end{equation}

    The $K$ value defines the $S$ distribution size. Consequently, it changes the coefficients $Q1$, $Q2$, $Q3$, and $IQR$ values that summarize $S$. As a result, it also causes changes in the local threshold value, $T_{related}$. Since $T_{related}$ is directly influenced by $S$, we call $T_{related}$ as `local threshold'.

\end{enumerate}

After setting the local threshold, the {\tt RD-Detector} detects the pairs of related discussion candidates. Related discussion candidates are those discussion pairs that similarity values are equal to or greater than the local threshold. {\tt RD-Detector} creates $R$, the set of related discussion candidates (Algorithm~\ref{algo:match}, lines 26-30). We consider related discussions those identified as outliers in the $S$ distribution. \cite{calefato2021will} also use descriptive statistics to identify core OSS developers' inactivity periods.

We measured the precision of the {\tt RD-Detector} based on the assessment of OSS maintainers and SE researchers who manually evaluated the $R$ set. Since we do not know the number of related posts or the truly related discussion posts in the \GHD forum, we focus on increasing the {\tt RD-Detector} precision rather than its recall. Higher precision values ensure greater assertiveness in detecting true positives.

The precision rate refers to the number of true positives divided by the approaches' total number of positive predictions. In this work, evaluators judged the {\tt RD-Detector} predictions and highlighted the true positives ones. We present the overall process of the {\tt RD-Detector} evaluation in Section~\ref{sec_caseSatudy_GH:evaluation}.

Let $N$ be the set of true-positive related discussions, and $R$ be the set of predicted related discussions made by {\tt RD-Detector}. The precision rate measurement is as follows \citep{kim2005improving}: 

\begin{equation}
\label{eq:precision}
  Precision = \frac{|N \cap R|}{|R|}
\end{equation}

The precision value ranges from 0\% to 100\%~\citep{kim2005improving}. A precision of 100\% means that all identified related discussion candidates are indeed related discussion posts.

The {\tt RD-Detector} emerges as a tool that OSS maintainers can use to minimize the work overhead in manually detecting related threads and reduce the rework in answering the same question multiple times. The periodic execution of {\tt RD-Detector} supports maintainers to deal with related discussions propagation in OSS communities.

\section{Assessing {\tt RD-Detector} over the \GHD forum}
\label{sec_caseSatudy_GH}

To assess {\tt RD-Detector}, we collected and used pairs of discussion posts of three selected OSS projects hosted on the GitHub platform (Table~\ref{tab:GH_repositories}). {\tt RD-Detector} generated sets of related discussion candidates for each project. Maintainers of the selected OSS projects and SE researchers evaluated the {\tt RD-Detector} outcomes.

 \subsection{Data collection - \GHD}
 \label{sec_caseSatudy_GH:dataset}

To perform the data collection, we wrote a customized web crawler that searches for discussion posts on the GitHub platform and collects all public discussion posts of a given OSS community. To do so, we used the Python BeautifulSoup library~\footnote{https://pypi.org/project/beautifulsoup4/}. We stored the collected data in a local dataset. The web crawler input is the project URL, and its outputs comprise of (1) a list with the discussions' IDs, categories, authors, dates, and title; and (2) the HTML files of the discussion threads (discussion first post and comments).

We collected data from {\tt Gatsby} ($D_{p=Gatsby}$), {\tt Homebrew} ($D_{p=Homebrew}$), and {\tt Next.js} ($D_{p=Next.js}$) repositories (Table~\ref{tab:GH_repositories}). We selected these three repositories based on (1) the usage of the Discussions forum, (2) the GitHub professional team members' demands (coauthors in this research), and (3) OSS maintainers' availability. We executed multiples data collection runs between October 2021 and November 2021. In total, we collected 11,309 discussions.

\begin{table}
    \centering
  \caption{Repositories used in the dataset}
  \label{tab:GH_repositories}
  \begin{tabular}{lcc}
    \toprule
    Repository&  \#Discussions & Download date\\

    \midrule

    {\tt Gatsby~\footnote{https://github.com/gatsbyjs/gatsby/discussions/}} & 1,276 & 11-10-2021\\
    {\tt Homebrew\footnote{https://github.com/homebrew/discussions/discussions/}} & 1,464 & 10-09-2021\\
    {\tt Next.js\footnote{https://github.com/vercel/Next.js/discussions/}} & 8,569 & 11-18-2021\\
    
    \midrule
    \textbf{sum} &\textbf{11,309}&\\
  \bottomrule
\end{tabular}
\end{table}
 
\subsubsection{Dataset characterization}
\label{sec_caseSatudy_GH:dtcharacterization}

\begin{figure}[h]

  \centering
  \includegraphics[width=\linewidth]{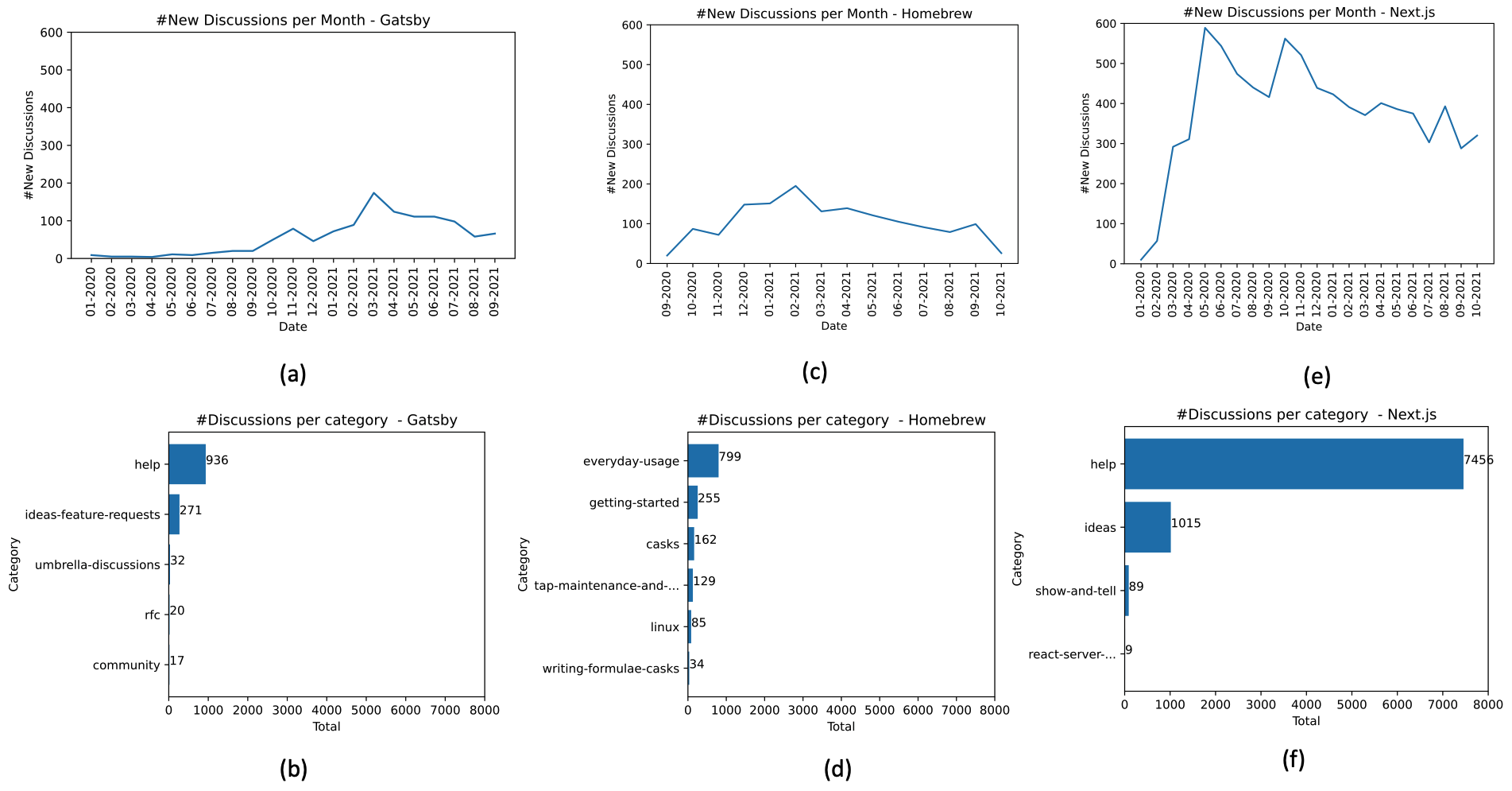}
  \caption{Dataset characterization - {\tt Gatsby}, {\tt Homebrew}, and {\tt Next.js} }

   \label{fig:dataset_characterization}
\end{figure}

According to the {\tt Gatsby} community documentation~\citep{gatsby_doc}, `Gatsby is a free and open source framework based on React that helps developers build blazing fast websites and apps.'. We call $D_{Gatsby}$ the set of discussion posts collected from {\tt Gatsby}'s repository. On the date we performed the data collection, we collected 1,276 discussion posts from {\tt Gatsby}, $|D_{p=Gatsby}| = 1,276$ (Table~\ref{tab:GH_repositories}). Figure~\ref{fig:dataset_characterization}-(a) shows the frequency distribution of discussions created in the {\tt Gatsby} community over 21 months (01-2020 to 09-2021). According to this time window, 01-2020 to 09-2021, the average growth rate of forum usage by the Gatsby community was 22\% (considering the frequency of new discussions).{\tt Gatsby} makes available the following categories types {\tt community}, {\tt help}, {\tt ideas-feature-requests}, {\tt RFC}, and {\tt umbrella-discussions}. {\tt Help} discussions are most common, totaling 73.35\%, followed by {\tt ideas-fearute-requests}, {\tt umbrella-discussions}, {\tt community}, and {\tt RFC}, totaling 21.23\%, 2.50\%, 1.33\%, 1.56\% of discussion posts, respectively (Figure~\ref{fig:dataset_characterization}-(b)).

{\tt Homebrew} is an \OSS project that makes it easy to `install the UNIX tools Apple didn't include with macOS. It can also install software not packaged for your Linux distribution to your home directory without requiring sudo.'\citep{homebrew_doc}. 
We call $D_{Homebrew}$ the set of discussion posts collected from {\tt Homebrew}'s repository. We collected 1,464 discussion posts from {\tt Homebrew}, $|D_{p=Homebrew}| = 1,464$. 

Figure~\ref{fig:dataset_characterization}-(c) shows the frequency distribution of discussion posts created in {\tt Homebrew} Discussions forum over 14 months (09-2020 to 10-2021). The oldest discussion post collected from {\tt Homebrew} dates from September 2020. According to data in $D_{Homebrew}$, the average growth rate of Discussions forum usage in {\tt Homebrew} between 09-2020 and 10-2021 was 25.12\% (considering the frequency of new discussions). The months with the highest frequency of new discussions were December/20, January, and February 2021, totaling 148, 151, and 195 new discussions, respectively. Such values indicate the peak in the frequency distribution graph (Figure~\ref{fig:dataset_characterization} (c)). Unlike {\tt Gatsby} and {\tt Next.js} projects, {\tt Homebrew} organizes its discussion posts according to the problem types and not by questions type. {\tt Homebrew} provides the following categories: {\tt casks}, {\tt getting-started}, {\tt tap-maintenance-and-brew-development}, {\tt everyday-usage}, {\tt linux}, and {\tt writing- \\formulae-casks}. The {\tt everyday-usage} discussions are the most common in the $D_{Homebrew}$ dataset, totaling 54.87\%. Followed by {\tt getting-started}, {\tt casks}, {\tt tap- maintenance-and-brew-development}, {\tt linux}, and {\tt writing-formulae-casks}, totaling 17.27\%, 10.85\%, 8.84\%, 5.80\% and 2.34\% of discussions, respectively (Figure~\ref{fig:dataset_characterization} (d)).

{\tt Next.js} is an OSS project that `provides a solution to build a complete web application with React'\citep{nextJS_doc}. 
The {\tt Next.js} community has stood apart in supporting the launch of \GHD since the start~\citep{Evi20}. In January 2022, the project forum already had almost 400 pages of discussion threads~\citep{Evi20}. Compared to {\tt Gatsby} and {\tt Homebrew}, {\tt Next.js} is the project with the highest number of discussions analyzed, characterizing an active OSS community. We collected 8,569 public discussion posts from {\tt Next.js}. Figure~\ref{fig:dataset_characterization} - (e) shows the frequency distribution of new discussions over 22 months (01-2020 to 10-2021). In this period, the average growth rate of the forum in the {\tt Next.js} community was 44.62\%(considering the number of new discussions created). The {\tt Next.js} project has the highest average growth rate among the three analyzed projects. The peaks of the distribution indicate the months with the highest number of discussions created, which were May, June, and October 2020 with 589, 544, and 562, respectively (Figure~\ref{fig:dataset_characterization} - (e)). When we collected the dataset, the discussion categories were {\tt ideas}, {\tt help}, {\tt react-server-components}, and {\tt show-and-tell}. {\tt Help} discussions are the most frequent in the {\tt Next.js} dataset, totaling 87.01\% of the discussion posts. Followed by {\tt ideas}, {\tt show-and-tell}, and {\tt react-server-components}, totaling 11.84\%, 1.03\% and 0.10\% of discussions, respectively (Figure~\ref{fig:dataset_characterization}-(f)).

Finally, all discussion posts collected composes the dataset $D$ used in this work. Therefore, $D=D_{p=Gatsby} \cup D_{p=Homebrew} \cup D_{p=Next.js}$. 

\subsection{Preprocessing phase applied to Discussions Dataset}
\label{sec_caseGH:prepocessing}

We preprocessed the dataset $D$. First, we split the dataset into subsets using predefined filters. We considered two filter types: project and category filters. We used the project filter, $p$, to select discussions from a specific OSS repository by defining the project's name. We used the category filter, $c$, to select discussion posts that matched predefined categories' types.

We set both filter values according to the dataset characterization. One can also combine filters to create different sub-datasets, e.g., we can select discussions of type $\beta$ belonging to project $\alpha$. The category filter values depend on the types of categories provided by the project. We express the use of both filters by $p=\alpha|c=\beta$, where $p$ identifies to the project name (ex.: {\tt Gatsby}, {\tt Homebrew}, and {\tt Next.js }) and $c$ identifies the discussion category type.

We standardized the category labels to report our results, making them easier to describe and follow. We use the following labels: {\tt Q\&A}, {\tt Ideas}, and {\tt ALL}. The {\tt Q\&A} label refers to `Question and Answers' (help) discussions types. The {\tt Ideas} category to refer to {\tt ideas} or {\tt ideas-feature-requests} categories. In addition, we defined a special label, {\tt ALL}, used to refer to all discussion posts, regardless of their original type. As described in Section~\ref{sec_caseSatudy_GH:dtcharacterization}, the {\tt Homebrew} project discussion organization differs from the other projects. Due to this fact, we chose not to filter {\tt Homebrew} discussions according to the category type. Therefore, we report the results of the {\tt Homebrew} project by setting $c=ALL$.

Besides, on the this step, {\tt RD-detector} selected the discussions' title and body text. We derived seven sub-datasets by applying different filter configurations to the original dataset $D$ (Table~\ref{tab:GH_datasets_filtros}, column 3). We use $D_{p=Gatsby|c=Q\&A}$ to designate the subset of Q\&A discussion collected from the {\tt Gatsby} project, $D_{p=Gatsby|c=Ideas}$ to denote the subset of {\tt ideas-feature-requests} discussions collected from the {\tt Gatsby} project, and so on. We assessed the proposed approach ({\tt RD-Detector}) over the seven sub-datasets. 

Next, the sub-datasets went through cleaning, denoising, and formatting steps. The last preprocessing step ignores zero-length discussions. In total, the approach discarded 3, 17, and 127 discussions from {\tt Gatsby}, {\tt Homebrew}, and {\tt Next.js} projects, respectively.

\subsection{Relatedness Checker applied to \GHD}
\label{sec_caseSatudy_GH:similarityChecker}

After the preprocessing phase, the {\tt RD-Detector} calculated the semantic similarity values for each pair of discussions (as presented on Section~\ref{sec_approach:similarityChecker}), using the seven sub-datasets presented in Table~\ref{tab:GH_datasets_filtros}, column 3. As discussed in Section~\ref{sec_approach:selection}, the value of $K$ impacts the set of related discussion candidates $R$. The {\tt RD-Detector} uses the $K$ value to select the similarity values of the top-K most similar discussions to each post in the dataset. In order to evaluate the {\tt RD-Detector} effectiveness, we set the $K$ value to 5 and 10. For each $D_{p=\alpha|c=\beta}$ sub-dataset presented in Table~\ref{tab:GH_datasets_filtros}, we configured the approach to run over $K=5$ and $K=10$. This way, we evaluate the approach's effectiveness by considering the five and the ten most similar discussions to every discussion in the dataset. Values greater than 10 expand the search bounds for related discussions. We did not consider $K$ values greater than ten because the proposed approach aims at increasing the precision rates rather than recall rates. 

In total, we assessed the approach over 14 different configuration groups (Table~\ref{tab:GH_datasets_filtros}, column 4). For each configuration group, {\tt RD-Detector} computed the local threshold value, $T_{related}$, and detected the set of related discussion candidates, $R$. We denote the set of related discussion candidates detected considering the configuration group $p=Gatsby$, $c=Q\&A$, and $K= 5$ by $R_{p=Gatsby|c=Q\&A|K=5}$. $R_{p=Gatsby|c=Q\&A|K=10}$ denotes the set of related discussion candidates detected considering the configurations $p=Gatsby|c=Q\&A|K=10$. One can use the same reasoning for the other 12 related discussion sets presented in Table~\ref{tab:GH_datasets_filtros}, column 4. We used the configuration groups to evaluate the {\tt RD-Detector} effectiveness and report our results. 

\begin{table}
\centering
  \caption{Configuration groups: projects, categories, and K-values}
  \label{tab:GH_datasets_filtros}
  \begin{tabular}{llll}
    \toprule
    \begin{tabular}[c]{@{}l@{}} Project \end{tabular}
     & 
   \begin{tabular}[c]{@{}l@{}} Category \\  \end{tabular}
 
   & Dataset & 
    
   \begin{tabular}[c]{@{}l@{}} Set of related discussion \\ candidates \end{tabular}
   
   \\
    
    \midrule
    
    \multirow{3}{*}{ {\tt Gatsby}}
          & {\tt Q\&A}

          & $D_{p=Gatsby|c=Q\&A}$ & 
          
          \begin{tabular}[c]{@{}l@{}} $R_{p=Gatsby|c=Q\&A|K=5}$
                                      \\ $R_{p=Gatsby|c=Q\&A|K=10}$
           \end{tabular}
          
          \\ 
        ~ & {\tt Idea} & $D_{p=Gatsby|c=idea}$ & 
        
        \begin{tabular}[c]{@{}l@{}} $R_{p=Gatsby|c=idea|K=5}$
                                    \\ $R_{p=Gatsby|c=idea|K=10}$
           \end{tabular}
        
        \\ 
        ~ & {\tt ALL} & $D_{p=Gatsby|c=ALL}$ &

        \begin{tabular}[c]{@{}l@{}} $R_{p=Gatsby|c=ALL|K=5}$
                                      \\ $R_{p=Gatsby|c=ALL|K=10}$
           \end{tabular}

        \\ 
        \hline
    
    \multirow{1}{*}{{\tt Homebrew}}
     & ALL &$D_{p=Homebrew|c=ALL}$ & 
     
     \begin{tabular}[c]{@{}l@{}} $R_{p=Homebrew|c=ALL|K=5}$
                                      \\ $R_{p=Homebrew|c=ALL|K=10}$
           \end{tabular}

     \\ 
        \hline
    
    \multirow{3}{*}{{\tt Next.js}}
         & {\tt Q\&A} & $D_{p=Next.js|c=Q\&A}$ &   
         \begin{tabular}[c]{@{}l@{}} $R_{p=Next.js|c=Q\&A|K=5}$
                                    \\ $R_{p=Next.js|c=Q\&A|K=10}$
           \end{tabular}
         
         \\ 
        ~ & {\tt Idea} & $D_{p=Next.js|c=idea}$ &   
        
        \begin{tabular}[c]{@{}l@{}} $R_{p=Next.js|c=idea|K=5}$
                                    \\ $R_{p=Next.js|c=idea|K=10}$
           \end{tabular}
        
        \\ 
        ~ & {\tt ALL} & $D_{p=Next.js|c=ALL}$ & 
         \begin{tabular}[c]{@{}l@{}} $R_{p=Next.js|c=ALL|K=5}$
                                    \\ $R_{p=Next.js|c=ALL|K=10}$
           \end{tabular}
        \\ 
    
  \bottomrule
\end{tabular}
\end{table}

\subsection{The RD-Detector evaluation}
\label{sec_caseSatudy_GH:evaluation}

We recruited maintainers of the three OSS projects analyzed and SE researchers to evaluate the {\tt RD-Detector} outcomes. Both OSS maintainers and SE researchers manually classified pairs of related discussion candidates as duplicates, related or not related. OSS maintainers judged pairs of related discussion candidates according to their work project. The SE researchers judged pairs of related discussion candidates detected on the three projects. The maintainers, M\_Gatsby, M\_Homebrew, and M\_Next.js, were contacted via the GitHub team of professionals. The three SE researchers, SE\_R1, SE\_R2, and SE\_R3, contacted have different expertise in software development. SE\_R1 is an industry practitioner and a Software Engineering researcher, SE\_R2 is an active OSS contributor and a Software Engineering researcher, and SE\_R3 is a Software Engineering researcher.

Evaluators received online documents containing instructions to evaluate the sets of the related discussion candidates. The documents (1) described the concept of duplicates and related discussions, and (2) listed pairs of related discussion candidates containing. We characterized each pair of related discussions by describing the ID and title of the master and target discussions. The ID was a link from which evaluators could access the original discussion posts. We instructed the evaluators to add the label `D'  to duplicates, `R' to related ones, and `N' for unrelated discussions. We also asked the evaluators to add comments to justify their judgments.

We measured the {\tt RD-Detector} precision rate compared to the OSS maintainers' and SE researchers' judgment. Their judgments identified the true-positive {\tt RD-Detector} predictions and defined the set $N$ used in Equation~\ref{eq:precision}. Section~\ref{sec_results} reports the precision rate reached assessing {\tt RD-Detector} over the 14 configurations groups presented in Table~\ref{tab:GH_datasets_filtros}. 

\section{Results}
\label{sec_results}

In this section, we first present the experimental results of {\tt RD-Detector} evaluation. Next, we describe the OSS maintainers' feedback about the detected related discussion candidates and the {\tt RD-Detector} practical applications.

To evaluate the effectiveness of our approaches, we ran {\tt RD-Detector} using different configurations (Table~\ref{tab:GH_datasets_filtros}) and measured its precision rate compared to the OSS maintainers' and SE researchers' judgment (Section~\ref{sec_caseSatudy_GH:evaluation}). First, to better analyze and report our results, we filtered the dataset according to the OSS repositories names and the discussion category. Next, we parameterized the approach to use different values of $K$. In total, combining different configurations, the {\tt RD-Detector} outputs 14 related discussion candidates sets (Table~\ref{tab:GH_datasets_filtros}), over which we assessed the {\tt RD-Detector} approach. The selected evaluators judged all 14 sets of related discussion candidates.

As discussed in Section~\ref{sec_approach:selection}, as one changes the $K$ value, the $T_{related}$ threshold also changes and, consequently, the number of detected related discussion candidate pairs. Tables~\ref{tab:descripStat_Gatsby}, \ref{tab:decripStat_Homebrew}, and \ref{tab:descripStat_NextJS} present the local threshold values, $T_{related}$,  and the number of related discussion candidates pairs detected, $|R|$, for {\tt Gatsby}, {\tt Homebrew}, and {\tt Next.js} projects, respectively. Tables~\ref{tab:descripStat_Gatsby}, \ref{tab:decripStat_Homebrew}, and \ref{tab:descripStat_NextJS} also show, for each configuration group (project name, category type, and $K$ value), the distribution size ($size(S)$) and the descriptive statistics coefficients that summarize $S$ ($IQR$, $Q1$, $Q2$, and $Q3$). We used the values of the coefficients to calculate the local threshold $T_{related}$ (Equation~\ref{eq:limiar}).

We can note that as we increase the value of $K$, the local threshold values, $T_{related}$, decrease (Tables~\ref{tab:descripStat_Gatsby}, \ref{tab:decripStat_Homebrew}, and \ref{tab:descripStat_NextJS}). Since the local threshold value decreases, the {\tt RD-Detector} detects new pairs of related discussion candidates. The new pairs are those outliers considered by changing the Upper Inner Fence value, according to Equations~\ref{eq:upper_inner_fence} and \ref{eq:limiar}.

\begin{table}[h]
	\centering
	\caption{Descriptive Statistics - {\tt Gatsby}}
	\label{tab:descripStat_Gatsby}
	{
		\begin{tabular}{lcccccc}
			\toprule
			 & \multicolumn{6}{c}{ {\tt Gatsby} - Similarity values } \\
			\cmidrule[0.4pt]{2-7}
			& \multicolumn{2}{c}{ c = Q\&A } &  \multicolumn{2}{c}{ c = Ideas } &
			\multicolumn{2}{c}{ c = ALL } \\
			 & $K = 5$ & $K = 10$  & $K = 5$ & $K = 10$ & $K = 5$ & $K = 10$\\
			 \cmidrule[0.4pt]{2-7}
			\textit{Size(S)} & 3924 & 7806 & 1063 & 2080 & 5277 & 10517 \\
			$IQR$ & 0.145 & 0.145 & 0.132 & 0.135 & 0.137 & 0.139\\
			$Q1$ & 0.566 & 0.542 & 0.499 & 0.461 & 0.571 & 0.545 \\
			$Q2$ & 0.648 & 0.622 & 0.568 & 0.540 & 0.649 & 0.623\\
			$Q3$ & 0.711 & 0.687 & 0.630 & 0.596 & 0.708 & 0.684 \\
			\cmidrule[0.4pt]{2-7}
			\textit{$T_{related}$} & 0.9278  & 0.9040 & 0.8278 & 0.7992 & 0.9127 & 0.8925 \\
			$|R|$ & 4 & 6  & 6 & 8 & 7 & 12\\
			\bottomrule
		\end{tabular}
	}
\end{table}

Table~\ref{tab:descripStat_Gatsby} presents the results for the Gatsby project. Considering $K=5$ and $c=Q\&A$, {\tt RD-Detector} calculated the local threshold value based on the similarity values of 3,924 unique discussion pairs, $size(S)$, setting $T_{related} = 0.9278$ and detecting four related discussion candidates, $|R|$. Considering the configuration group $|p=Gatsby|c=Q\&A|K=10$, the local threshold value, $T_{related} = 0.9040$, was calculated based on 7,806 similarity values of unique pairs of discussions, detecting six pairs of related discussion candidates. Besides, considering $c=Ideas$ and $K=5$, the local threshold $T_{related}$ calculated was 0.8278. This value was based on the analysis of 1,063 similarity values and detected six pairs of related discussion candidates. In addition, for the configuration group $p={\tt Gatsby}|c=Ideas|K=10$, the {\tt RD-Detector} detected eight pairs of related discussion candidates, considering $T_{related} = 0.7992$. Finally, one can apply the same standard interpretation to Gatsby's results by fixing $c=ALL$ and varying $K=5$ and $K=10$.

\begin{table}[h]
	\centering
	\caption{Descriptive Statistics - { \tt Homebrew} }
	\label{tab:decripStat_Homebrew}
	{
		\begin{tabular}{lcc}
			\toprule
			\multicolumn{3}{c}{ {\tt Homebrew} - Similarity values }\\
			\cmidrule[0.4pt]{2-3}
			
			&\multicolumn{2}{c}{ c = ALL } \\
			& $K = 5$ & $K = 10$\\
			\cmidrule[0.4pt]{2-3}
			\textit{Size(S)} & 6009 & 11948 \\
			$IQR$ & 0.130 & 0.130\\
			$Q1$ & 0.535 & 0.508  \\
			$Q2$ & 0.602 & 0.578 \\
			$Q3$ & 0.665 & 0.638\\
			\cmidrule[0.4pt]{2-3}
			\textit{$T_{related}$} & 0.8592 & 0.8336  \\
			$|R|$ & 15 & 34  \\
			\bottomrule
		\end{tabular}
	}
\end{table}

Table~\ref{tab:decripStat_Homebrew} presents the {\tt RD-Detector} outcomes for the {\tt Homebrew} project. Considering the configuration group $p={\tt Homebrew}|c=ALL|K=5$, the {\tt RD-Detector} calculated the local threshold value based on the similarity values of 6,009 unique discussion pairs, $size(S)$, achieving $T_{related} = 0.8592$. As we can see in Table~\ref{tab:decripStat_Homebrew}, the approach identified 15 pairs of related discussion candidates considering $c=ALL$ and $K=5$. However, considering $c=ALL$ and $K=10$, the approach calculated the local threshold value, $T_{related} = 0.8336$, based on 11,948 similarity values. Using this last configuration group, the {\tt RD-Detector} detected 34 related discussion candidate pairs.

\begin{table}[h]
	\centering
	\caption{Descriptive Statistics - { \tt Next.js} }
	\label{tab:descripStat_NextJS}
	{
		\begin{tabular}{lcccccc}
			\toprule
			&\multicolumn{6}{c}{ {\tt Next.js} - Similarity values }\\
			\cmidrule[0.4pt]{2-7}
			& \multicolumn{2}{c}{c = Q\&A } &  \multicolumn{2}{c}{c = Ideas } &
			\multicolumn{2}{c}{c = ALL } \\
			 & $K = 5$ & $K = 10$  & $K = 5$ & $K = 10$ & $K = 5$ & $K = 10$\\
			\cmidrule[0.4pt]{2-7}
			\textit{Size(S)} & 30665 & 60704 & 4006 & 7834 & 35171 & 69616 \\
			$IQR$ & 0.111 & 0.113 & 0.119 &  0.114 & 0.109 & 0.111\\
			$Q1$  & 0.581 & 0.558 & 0.543 & 0.514 & 0.588 & 0.565  \\
			$Q2$  & 0.639 & 0.617 & 0.607 & 0.574 & 0.646 & 0.623 \\
			$Q3$  & 0.693 & 0.672 & 0.662 & 0.629 & 0.697 & 0.676\\
			\cmidrule[0.4pt]{2-7}
			\textit{$T_{related}$} & 0.8605  & 0.8427 & 0.8402 & 0.8014& 0.8608 & 0.8441\\
			$|R|$ & 95 & 122  & 10 & 35 & 106& 137\\
			\bottomrule
		\end{tabular}
	}
\end{table}

Table~\ref{tab:descripStat_NextJS} presents the distribution size, $Size(S)$, and the number of detected related discussion pairs, $|R|$,  for the {\tt Next.js} project. These numbers endorse the scenario described by \cite{ahasanuzzaman2016mining}. The authors highlighted that as PCQA forums become popular, the number of posts increases. Consequently, it may also increase the number of duplicates or related posts. Since the {\tt Next.js} project is the largest considering the number of discussion posts (Table~\ref{tab:GH_repositories}),  the chances of related discussion occurring can increase. Using the configuration group $|p={\tt Next.js}|c=ALL|K=10$, the {\tt RD-Detector} detected 137 pairs of related discussion candidates - the most extensive related discussion set detected. For this last configuration group, the local threshold value, $T_{related}=0.8441$, was calculated based on the similarity value of 69,616 unique pairs of discussions (Table~\ref{tab:descripStat_NextJS}).

For each of the 14 configuration groups presented in Table~\ref{tab:GH_datasets_filtros} column 4, the {\tt RD-Detector} calculated different values of local thresholds and detected different amounts of related discussion candidate pairs. We highlight that all sets of related discussion candidates detected considering $K=5$ are subsets of the related discussion candidate sets setting $K=10$. Table~\ref{tab:result_precision} shows the precision rate reached by the {\tt RD-Detector} considering the 14 configuration groups.

\begin{table}[h!]
	\centering
  \caption{Evaluation Results}
  \label{tab:result_precision}
  \begin{tabular}{lcc}
  \toprule
  
  Category & $K$ = 5 & $K$ = 10 \\
  \midrule
  
  & \multicolumn{2}{c}{{\tt Gatsby}} \\
  \midrule
  
    Q\&A  &  100\% &  100\%   \\
    Ideas  &  83.33\% & 75\%  \\
    ALL  &  100\%  & 91.66\%  \\
  \toprule
  &\multicolumn{2}{c}{{\tt Homebrew}} \\
  \midrule
    ALL  &  100\%  & 91.17\%  \\
   \toprule
   &\multicolumn{2}{c}{{\tt Next.js}} \\
   \midrule
    Q\&A  &  95.78\%  & 90.98\%  \\
    Ideas  &  90\%   & 88.57\%  \\
    ALL  &  97.16\%  & 93.43\%   \\
  \bottomrule 
\end{tabular}
\end{table}

The smallest set size of related discussion candidates was detected using the configuration group $p={\tt Gatsby}|c=Q\&A|K=5$ (Tables~\ref{tab:descripStat_Gatsby}, \ref{tab:decripStat_Homebrew}, and \ref{tab:descripStat_NextJS}). According to this configuration group, the approach detected four related discussion candidate pairs, $|R_{p={\tt Gatsby}|c=Q\&A|K=5}| = 4$. Evaluators emphasized that all four pairs are truly related discussions, which means that all of them contain discussion posts related to one other. On the other hand, the {\tt RD-Detector} approach detected the biggest set of related discussion candidates using the configuration $|p={\tt Next.js}|c=ALL|K=10$. In total, the approach detected 137 pairs of related discussion candidates,  $|R_{p={\tt Next.js}|c=ALL|K=10}| = 137$, of which 128 were classified as related by the evaluators, achieving a precision rate of 93.43\% (Table~\ref{tab:result_precision}).

The OSS maintainer, M\_Gatsby, and two SE researchers evaluated {\tt Gatsby} project's related discussion candidates. In order to save M\_Gatsby's time and effort in judging cases of duplicates that are exact copies, the GitHub professional team requested researchers to judge pairs of related discussion with a high similarity value. Two SE researchers judged pairs of related discussion candidates with similarity values greater than or equal to 0.9415. The researchers agreed that all judged pairs contained related discussions. The maintainer M\_Gatsby judged related discussion candidate pairs that required prior technical knowledge about the project. According to Table~\ref{tab:result_precision}, {\tt RD-Detector} reached the maximum precision value (100\%) in detecting related discussion pairs fixing $c=Q\&A$ and varying the $K$ value ($K=5$ and $K=10$) for the {\tt Gatsby} project. Considering the {\tt Gatsby} project discussions classified as Ideas ($c=Ideas$), the approach achieved an 83.33\% precision rate for $K=5$ and 75\% setting $K=10$. Still processing {\tt Gatsby}'s discussions and setting $c=ALL$, the {\tt RD-Detector} achieved a precision rate of 100\% and 91.66\%  for $K=5$ and $K=10$, respectively.

The maintainer M\_Homebrew and two SE researchers evaluated the {\tt Homebrew} project's set of related discussion candidates. As we already mentioned, to save the maintainer time and effort, the researchers judged pairs of related discussion candidates with the highest similarity value (greater than or equal to 0.9558). Considering the set $R_{p={\tt Homebrew},c=ALL,K=5}$, the {\tt RD-Detector} achieved the highest precision rate in detecting related discussion pairs. Evaluators judged all of the 15 pairs of related discussion candidates presented in $R_{p={\tt Homebrew},c=ALL,K=5}$ as related. However, setting the configuration group $|p={\tt Homebrew}|c=ALL|K=10$, the approach achieved 91.17\% of the precision rate. From evaluators' perspective, 31 out of 34 are truly related pairs of discussion posts. The maintainer M\_Homebrew judged the discussions of three pairs as unrelated.

As discussed in Section~\ref{sec_caseSatudy_GH:dtcharacterization}, compared to {\tt Gatsby} and {\tt Next.js} projects, the {\tt Homebrew} project organized the discussion posts differently. It does not provide the same default categories provided by {\tt Gatsby} and {\tt Next.js}. {\tt Homebrew} discussions categorization is according to the problem type. That is why we did not split the {\tt Homebrew} dataset by category and  present the achieved precision rate setting the category filter to $ALL$ ($c=ALL$).

Finally, the number of detected related discussion candidates for the {\tt Next.js} project required greater participation of SE researchers in the evaluation phase. Because the related discussion candidate sets of the {\tt Next.js} project are the largest, the project maintainer judgment became unfeasible. To ensure the reliability of the reported results, we measured the researchers' inter-rater agreement using the Cohens Kappa Coefficient~\citep{cohen1960coefficient}, which was 0.85. This value indicates almost perfect agreement according to the interpretation proposed by \cite{landis1977measurement}. Setting the configuration group $|p=Next.js|c=Q\&A|K=5$, the approach reached precision values of 95.78\%. The {\tt RD-Detector} reached a precision of 90.98\%, setting the $K$ value to 10. Fixing $p={\tt Next.js}$ and $c=Ideas$, the {\tt RD-Detector} reached precision rates of 90\% and 88.57\% for $K=5$ and $K=10$, respectively. Besides, when we fixed $p={\tt Next.js}$ and $c=ALL$, the {\tt RD-Detector} reached precision rates of 97.16\% and 93.43\% for $K=5$ and $K=10$, respectively.

Table~\ref{tab:result_precision} shows that precision rates degrade as we increase the value $K$. The {\tt RD-Detector} identified 18, 34, and 166 unique pairs of related discussion candidates for the {\tt Gatsby}, {\tt Homebrew}, and {\tt Next.js} projects, respectively. The different numbers of detected related discussion pairs highlight the uniqueness of the OSS communities regarding the use of the \GHD forum. It also endorses the need for approaches that meet local thresholds to detect related discussion pairs and adapt to different contexts of OSS communities.

Table~\ref{tab:results_FP} lists all discussion pairs judged unrelated, grouped by project. The table presents the master and target discussions identifiers (IDs) and titles and their respective similarity values. Regarding {\tt Gatsby} project results, two pairs of related discussion candidates were classified as unrelated by the M\_Gatsby maintainer. Thus, 16 out of 18 detected pairs were indeed related (precision=88.88\%). From the perspective of the maintainer M\_Homebrew, out of the 34 related discussion candidate pairs detected, three are false positives. Therefore, 91.17\% of the detected pairs are truly related discussions. Finally, 90.96\% of discussion pairs detected for the {\tt Next.js} project are truly related. Out of 166 detected pairs, 15 are false positive. We discuss the false positives in Section~\ref{sec_discussion}.\\

\begin{table}
    
    \caption{False-positive related discussion candidates.}
    \label{tab:results_FP}
    \centering
    \setlength{\tabcolsep}{1pt}
    
    \resizebox{!}{5cm}{
    
    {\scriptsize

    \begin{tabular}{cclclc}
    \hline
        \# 
        & \begin{tabular}[c]{@{}c@{}} ID \\ Master \end{tabular}
        & \begin{tabular}[c]{@{}c@{}} Title Master \\ \end{tabular}
        & \begin{tabular}[c]{@{}c@{}} ID \\ Target \end{tabular}
        & \begin{tabular}[c]{@{}c@{}} Title Target \\ \end{tabular} 
        & \begin{tabular}[c]{@{}c@{}} Similarity \\ value \end{tabular}\\ \hline

        \toprule
        ~ & \multicolumn{4}{c}{{ \tt Gatsby}} \\
        \hline
        
        1 & 29766 & 
        
        \begin{tabular}[c]{@{}l@{}} I want to lint ts file with Gatsby's native \\ support. \end{tabular}
        
        & 32122 &

        \begin{tabular}[c]{@{}l@{}} On creating a better method to extend the \\ default ESLint configuration \end{tabular}
        
        & 0,897 \\ \hline
        2 & 26984 & 
        
        \begin{tabular}[c]{@{}l@{}} More flexible default Typescript transpiler \\ handling  \end{tabular}
        
        & 31662 & 
        
        \begin{tabular}[c]{@{}l@{}} Need better CSS modules + \\ Typescript support  \end{tabular}

         & 0,8172 \\ \hline

         \toprule
        ~ & \multicolumn{4}{c}{{ \tt Homebrew}} \\
        \hline

        3 & 1906 & 
        \begin{tabular}[c]{@{}l@{}} Questions on Homebrew's third-party \\ mirroring policy  \end{tabular}
        
        & 1917 & Setting up mirrors for Homebrew bottles & 0,8528 \\ \hline
        4 & 814 & 
        \begin{tabular}[c]{@{}l@{}}
        
        Homebrew not installing on Big Sur v 11.2.1\\ - Failed during: git fetch --force origin
        
         \end{tabular}
        
         & 1218 & Can't get homebrew installed & 0,8396 \\ \hline
        5 & 72 & Installing Homebrew on a Windows OS & 657 & Unable to install HomeBrew & 0,8353 \\ \hline

         \toprule
        ~ & \multicolumn{4}{c}{{ \tt Next.js}} \\
        \hline
    
        6 & 13368 
        & 
        
        \begin{tabular}[c]{@{}l@{}} Can I replace next-routes with \\ the new versions?  \end{tabular}

        & 20361 & nextjs with next-routes upgrade & 0,9281 \\ \hline
        7 & 15780 & 
        \begin{tabular}[c]{@{}l@{}}
        ModuleNotFoundError: Module not found: \\Error: Can't resolve 'fs' 
        \end{tabular}

        & 19154 & import sub module of other module is fail & 0,8887 \\ \hline
        8 & 13617 
        &   
        
        \begin{tabular}[c]{@{}l@{}} NextJS not showing TypeScript  \\ errors on Runtime  \end{tabular}

        & 24996 & 
        
        \begin{tabular}[c]{@{}l@{}} I am unable to setup ``typescript-is'' \\ with nextjs.  \end{tabular}

        & 0,8646 \\ \hline
        
        9 & 23195 
        &

        \begin{tabular}[c]{@{}l@{}} Tailwind CSS not being bundled in \\ static export   \end{tabular}
        
        & 25845 &

        \begin{tabular}[c]{@{}l@{}} NextJS with Tailwind does not work when \\ importing from Global.css \end{tabular}
         & 0,8624 \\ \hline
        10 & 23871 & 
        
        \begin{tabular}[c]{@{}l@{}}
        Typescript - Can you change where Next.js \\ outputs the "next-env.d.ts" file?
        \end{tabular}
        
          & 24996 & 
          
          \begin{tabular}[c]{@{}l@{}} I am unable to setup ``typescript-is'' \\ with nextjs.  \end{tabular}

          & 0,8593 \\ \hline
        11 & 14416 & deploy to Vercel error(Build error occurred) & 21107 & 
        
        \begin{tabular}[c]{@{}l@{}}
        Deploy Vercel Error 'Error: Command \\ "yarn run build" exited with 1'  
        \end{tabular}
        
        & 0,8522 \\ \hline
        12 & 19089 & 
        
        \begin{tabular}[c]{@{}l@{}}
        Image Component with AWS Throws 400 \\ Errors in Production
        \end{tabular}
         & 19568 &

         \begin{tabular}[c]{@{}l@{}} Image component not work when next is \\ deployed in sub path \end{tabular}
         
          & 0,8456 \\ \hline
        13 & 24778 & Cannot process DELETE request via CORS  & 29589 
        
        &  
        
         \begin{tabular}[c]{@{}l@{}}same-origin policy on API routes not \\ working \end{tabular}
        
        & 0,8455 \\ \hline
        14 & 15613 & 
         \begin{tabular}[c]{@{}l@{}}
         
        next export - missing stylesheets - sass \\files not being linked in the static html export 
         \end{tabular}
         & 23945 & 
         
         \begin{tabular}[c]{@{}l@{}} [Sass] Errors in basic use of Sass with \\ 3rd party node modules \end{tabular}
         
          & 0,8449 \\ \hline
        15 & 14508 & Props are undefined from getServerSideProps & 20529 
        
        &   
        
        \begin{tabular}[c]{@{}l@{}} Run getServerSideProps again on back \\ broswer \end{tabular}

        & 0,8446 \\ \hline
        16 & 22688 &

        \begin{tabular}[c]{@{}l@{}} Changes to .env.local are not loaded even \\ after restarting
        \end{tabular}
        
         & 22918 & Problem with less loader in next@10.0.8 & 0,843 \\ \hline
        17 & 17320 & [RFC] ESLint out of the box  & 24900 & ESLint in Next.js and Create Next App & 0,8479 \\ \hline
        18 & 17320 & [RFC] ESLint out of the box  & 28571 & Thoughts on Conformance in Next.js & 0,8099 \\ \hline
        19 & 24850 & Support Promises in next.config.js & 24851 &
        
        \begin{tabular}[c]{@{}l@{}} (re)Expose `nextConfig` on top-level \\ NextServer for custom servers  \end{tabular}
        
        & 0,8088 \\ \hline
        20 & 24850 & Support Promises in next.config.js & 31205 & runtime env vars in next.config.js & 0,8037 \\ \hline
    \end{tabular}
    }
    }
\end{table}

\textbf{Answering our RQ:} 
\textbf{The general purpose Sentence-BERT model applicable to NLP problems effectively detects related discussion posts held in the \GHD forum. The results presented in Table~\ref{tab:result_precision} show the effectiveness of using the {\tt all-mpnet-base-v2} model to compute the sentence embeddings of the discussion pots and detect related discussions.}

\subsection{OSS maintainers' perspective regarding the judgment of related discussion candidates}
\label{sec_results:mainteiners_perspective}

We asked OSS maintainers to comment on their judgment regarding their decision about the related discussion candidates.
Based on their comments, we could identify (1) challenges deciding what are duplicate and related discussions, (2) reasons why users create related discussions, and (3) practical applications of {\tt RD-Detector}.

    \subsubsection{Challenges deciding what are duplicate and related discussions}

We noted that the concept of `related discussions' --- and even `duplicate discussions' --- depends on the evaluators' perspective. Therefore, we see this conceptual imprecision as a limitation that may introduce some bias on the {\tt RD-Detector} evaluation. 

Based on maintainers' feedback regarding the sample of related discussion candidates analyzed, we found four ways to conceptualize duplicates:

\begin{enumerate}
    \item \textit{Duplicates can be exact copies} - `...(the discussions) are identical duplicates...' (M\_Next.js).
    
    \item \textit{Duplicates address the same issues}  - `The initial problem is essentially the same and they have the same primary solution...' (M\_Homebrew), `The two discussions are about the same issue and the solution is functionally the same, although the specific commands given are slightly different...' (M\_Homebrew), `The problem the user in the target discussion is experiencing appears to be identical (as noted by them in their description)...' (M\_Homebrew), `These topics are identical ...' (M\_Homebrew), `Different context but same problem and solution.'(M\_Next.js).

   \item \textit{Duplicates share common information (topic overlapping)} - `...the target discussion also discusses a secondary problem that was raised during the initial fix.' (M\_Homebrew), `Same thing exposed using an external library for data fetching' (M\_Next.js).

    \item \textit{Sometimes, the same user can create duplicate discussions} - `...and were opened by the same user' (M\_Homebrew), `Posted by the same author also.' (M\_Next.js).

\end{enumerate}

We found that the conceptualization of duplicates tends to vary according to objective and subjective criteria. The first criterion - exact copies - is an objective one. To judge duplicates that are exact copies does not require evaluators to do a deep semantic analysis of the discussions' content. Exactly copies tend to have high similarity values (the similarity value of the discussion pair judged by M\_Next.js as duplicates is 0.9793). However, the second and the third criteria require evaluators to analyze discussions semantically in order to identify the overlapping topic occurrence. Although the fourth criterion is objective, evaluators can not consider it separately to judge duplicates. 

Compared to duplicates, maintainers classified related discussions as those posts that are not precisely copies but have similar topics and share similar information. They justify their judgment by highlighting that related discussions:

\begin{enumerate}
    \item \textit{Are not precisely copies} - `...similar problems but not an exact duplicate' (M\_Homebrew), `Since different users are on different networks they will never be 100\% duplicate.' (M\_Homebrew), `These are all related to images so I'd say they're 'related', they're not exactly the same though' (M\_Next.js).
    
    \item \textit{Have the same solution and similar problems} - `Same solution and similar problems...' (M\_Homebrew), `These (discussions) are both discussing the same thing...' (M\_Homebrew), `Both users are experiencing the same initial problem...' (M\_Homebrew), `...These are all related to images...' (M\_Next.js).
    
    \item \textit{Express complimentary topics} - `The target discussion is actually a continuation of the main discussion...' (M\_Homebrew), `This one is the initial RFC of a feature implementation, the other discussion is a post release issue.' (M\_Next.js), `This discussion caused the 14890 RFC to be created' (M\_Next.js).
    
    \item \textit{Cover different aspects of the same topic} - `Very similar issue on the surface that relates to the same initial problem, but the actualy underlying problem in the two was totally different' (M\_Homebrew), `...One user opened the discussion for help because the suggested solution didn't work. The other user is asking why the problem occured in the first place.' (M\_Homebrew), `A feature request for the image optimization feature, related but talking about different things.' (M\_Next.js), `So the overall topic (guide) is the same, the guide which it should be about is different.' (M\_Gatsby).
    
    \item \textit{May address the same problem in different contexts} - `...is the same exact bug... but different context.' (M\_Next.js).

\end{enumerate}

Given the subjective conceptualization of both duplicates and related discussions concepts, and based on OSS maintainers' feedback, we note they also faced challenges in classifying the discussion pairs as duplicates or related - `This blurs the lines a bit between R and D...' (M\_Homebrew), `There's definitely some similarity between the two, but it's not really clear whether the issue is the same...So, it's hard to tell, but I don't think they're the same' (M\_Homebrew),  `...I would consider them to be duplicates without the additional context of what was needed to fix the issue in the target discussion' (M\_Homebrew), `I'm struggling with this one bit. They're technically different questions, so I understand why the user opened both (they both have the same author)...' (M\_Homebrew), `Looks very similar, almost duplicated ... but he has a solution using a different...' (M\_Next.js). This challenge endorses our decision to consider duplicates as a subtype of related discussions.

We also note that the way users describe their questions, problems, or requests may impact the {\tt RD-Detector} results and the evaluator's judgment. Sometimes, it can be challenging for users to state the issue clearly and with enough contextual detail: `Both people had a similar problem but described so vaguely that it could be anything' (M\_Homebrew).

    \subsubsection{Reasons to create related discussions}

Based on the feedback of {\tt Homebrew}'s maintainer, we note the three reasons Discussions users create related discussion posts. Although we likely do not list all causes, the following three items enable OSS maintainers to understand the reasons behind the occurrence of related discussions and plan actions to deal with them. Users of \GHD create related discussions because:

\begin{enumerate}
    \item \textit{They want to emphasize their need for help} - `...  After the maintainers stopped responding, the user opened a new discussion in an attempt to continue.' (M\_Homebrew).
    \item \textit{They want to add new information to better detail the discussion's main topic} - `The user never got a response, so they opened a new issue with some more information a few weeks later.' (M\_Homebrew).
    \item \textit{They want to ask for different solutions to the same issue} - `...but they are unhappy with the solution presented in the master discussion and would like an alternative' (M\_Homebrew).

\end{enumerate}

Specifically, for the {\tt Homebrew} project results, the maintainer identified multiple pairs of related discussions related. The maintainer states that the discussion posts refer to a peculiar situation, `These are a little tricky... .' Although they are related discussions, `so, I'd call the issues duplicates because they all are the same problem...', the solutions of the problems depend on the users' hardware and software configuration, `..each situation is different it totally depends on the user's specific machine and configuration.' The maintainer further states that the solutions are different and that there is no single way to help users, `The actual solutions may be different, but there's no way for us to help with that so...' Consequently, they provide the same cues to users participating in the multi-related discussions `...from our perspective, the advice we can give is the same' (M\_Homebrew). Based on this feedback, we envision the {\tt RD-Detector} can also provide results to maintainers prioritize project issues solving or intervene in recurring situations demanded by the OSS communities.

    \subsubsection{Practical applications of {\tt RD-Detector}}

Finally, maintainers suggested some practical applications for {\tt RD-Detector} results, as follows :

\begin{enumerate}
    \item \textit{to combine the discussions' content merging the related discussion threads} - `These (discussions) could have been combined into one discussion and it would have made sense...' (M\_Homebrew), `... But, if it were up to me, they should have gone together in the same discussion.' (M\_Homebrew).

    \item \textit{to move a discussion content to another location reorganizing the discussions threads as comments to each other} - `...the new issue should probably have been posted as a comment in the master discussion' (M\_Homebrew), `...(they) would've received better traction as a comment on one another.' (M\_Next.js), `This discussion could've sufficed as a comment on 21633.' (M\_Next.js).

    \item \textit{to recruit collaborators for specific tasks} - `...could be useful though for people looking for other guides to contribute to (in this instance).' (M\_Gatsby).

\end{enumerate}
  
The variety of OSS projects on the GitHub platform provides opportunities to develop innovative approaches to detect related discussion candidates on GitHub Discussions. However, the significant number of OSS communities, the projects' singularity, and the mentioned limitations challenge developing and validating such systems. Those reasons endorse the {\tt RD-Detector} design decision to use a general-purpose machine learning model, calculate local threshold values, and achieve better precision rates.

\section{Discussions}
\label{sec_discussion}

As we reported in Section~\ref{sec_results}, increasing the $K$ values impacts on the local threshold values and increases the number of detected related discussion candidates.
In this section, we discuss the effects of changing the $K$ value, the false positives presented in Table~\ref{tab:results_FP}, and the implications of this research from the perspective of the OSS communities and software engineering researchers.

\subsection{The impacts of changing the $K$ value}

As discussed in Section~\ref{sec_approach:selection}, the $K$ value delimits the search bounds for related discussion candidates. As one changes the value of $K$, the threshold $T_{related}$ also varies and, consequently, the number of detected related discussion candidates. By configuring the {\tt RD-Detector} to run over $K=5$ and $K=10$, we note that the sets of related discussion candidates created when we set $K=5$ are subsets of $K=10$. Consequently, there is a risk of false-positive predictions propagation through the related discussion candidates' sets. We identified this propagation problem by analyzing the sets of related discussion candidates detected considering the configuration groups $p=Gatsby$ and $c=Ideas$. The same unrelated discussion pair $(master, target)$ occurs in $R_{p={\tt Gatsby}|c=Ideas|K=5}$ and $R_{p={\tt Gatsby}|c=Ideas|K=10}$. Although the precision rate tends to decrease, the approach detects new pairs of related discussions when we vary the $K$ values from 5 to 10.

\textit{Gatsby project}: Analyzing the sets of candidates $R_{p={\tt Gatsby}|c=Q\&A|K=5}$ and $R_{p={\tt Gatsby}|c=Q\&A|K=10}$, {\tt RD-Detector} detected two new pairs of related discussions when changing the $K$ value from 5 to 10. The maintainer of the {\tt Gatsby} project, M\_Gatsby, judged discussions from both new pairs as related. Since all related discussion candidates from both sets were indeed related, the {\tt RD-Detector} achieved the best precision rate (100\%). Considering $p={\tt Gatsby}|c=Ideas|K=5$ and $p={\tt Gatsby}|c=Ideas|K=10$, the {\tt RD-Detector} also detected two new pairs of related discussion candidates by increasing the $K$ value. However, one out of the two new pairs was judged unrelated by M\_Gatsby. In this case, the precision rate decreased from 83.33\% to 75\%.

As discussed in Section~\ref{sec_caseGH:prepocessing}, when we set the category filter to {\tt ALL} ($c=ALL$), the {\tt RD-Detector} calculates the similarity values between all discussion pairs in the dataset. The approach detected seven related discussion candidates using the configuration $p=Gatsby|c=ALL|K=5$ ($R_{p={\tt Gatsby}|c= ALL|K=5}=7$), of which five elements are also present in set $R_{p={\tt Gatsby}|c= Q\&A|K=10}$. Evaluators judged all five elements in $R_{p={\tt Gatsby}|c= Q\&A|K=10}$ set as related. M\_Gatsby judged that the discussions of the two new pairs were related. We analyzed the discussions' content and found that users create related discussions in different categories. The discussions of these two specific pairs were from {\tt Idea} and {\tt Q\&A} categories, respectively.

Regarding the set $R_{p={\tt Gatsby}|c=ALL|K=10}$, six of the 12 pairs are also present in the set $R_{p={\tt Gatsby}|c=Q\&A|K=10}$, two pairs are in $R_{p={\tt Gatsby}|c=Idea|K=5}$, and two pairs also belongs to $R_{p={\tt Gatsby}|c=ALL|K=5}$. All those ten related discussion pairs contain indeed related discussion posts. The maintainer endorsed the relatedness between the discussion posts of one new pair and refuted the second one. By increasing the value of $K$ ($K=5$ to $K=10$) and setting $p={\tt Gatsby}|c=ALL$, the {\tt RD-Detector} detected five new pairs of related discussion candidates, of which one was judged as unrelated by M\_Gatsby.

\textit{Homebrew project}: Analyzing the sets of related discussion candidates created considering the configuration groups $p={\tt Homebrew}|c=ALL|K=5$ and $p={\tt Homebrew}|c=ALL|K=10$, the {\tt RD-Detector} detected 19 new related discussion candidates when changing the $K$ value from 5 to 10. Out of the new 19 candidates, M\_Homebrew judged three pairs as not related (Table~\ref{tab:results_FP}). The precision rate decreased from 100\%, for $p={\tt Homebrew}|c=ALL|K=5$, to 91.17\% for $p={\tt Homebrew}|c=ALL|K=10$.

\textit{Next.js project}: The {\tt RD-Detector} detected 27 new related discussion candidates when we varied the $K$ value and fixed $p=Next.js|c=Q\&A$. Out of the 27 new candidates, SE researchers judged seven pairs as not related. This means that the approach detected 20 new pairs of related discussions by increasing the value of $K$. However, the false positives decreased the {\tt RD-Detector} precision rate from almost 96\% to nearly 91\%. This scenario repeats to $p=Next.js|c=Ideas$ and $p=Next.js|c=ALL$. When we changed the $K$ value from 5 to 10, the approach detected 25 and 31 new related discussion candidates for $c=Ideas$ and $c=ALL$, respectively. In total, three and six new candidates for $c=Ideas$ and $c=ALL$ were judged unrelated by SE researchers, respectively. In both cases, the precision rate decreased.

Because of the intersection relationship between sets of related discussion candidates, SE researchers had already judged 101 of the 106 pairs of related discussion candidates in set $R_{p={\tt Next.js}|c=ALL|K=5}$. Researchers evaluated the new five pairs as related. The five pairs have two common characteristics: (1) they had discussions created by the same user, and (2) they had one of the discussions created as {\tt Q\&A} and the other as {\tt Ideas}. This finding corroborates the {\tt Gatsby} project findings. Users create related discussion posts in different categories. Regarding $R_{p={\tt Next.js}|c=ALL|K=10}$, researchers had already judged 133 out of the 137 related discussion pairs in the set. They also judged the four new pairs as being related.

Maintainers can set the value of $K$ according to their respective interests. Decreasing the $K$ value increases the {\tt RD-Detector} precision rate. Higher precision values ensure greater assertiveness in detecting true positives. Conversely, increasing the $K$ value may reduce the precision rate. However, increasing the $K$ value also increases the number of detected related discussion candidates.

\subsection{False-positive {\tt RD-Detector} predictions}

Four authors of this paper manually analyzed the false positives presented in Table~\ref{tab:results_FP}. Based on evidence extracted from the discussion posts, we identified some limitations on using the proposed approach. We describe the false positive as follows.

\textit{Gatsby project}: The two false-positive cases are related to discussions classified as $Idea$. The researchers identified that the {\tt RD-Detector} did not capture the project-related specifics on both pairs. Although the two posts of pair \#1 (Table~\ref{tab:results_FP}) address the same topic (`JavaScript linting utility ESLint') and have keywords intersection, they address different problems. The discussion posts of the \#2 pair (Table~\ref{tab:results_FP}) address the topic `Typescript' and have project keywords intersection. However, the specificity of the issues described in the discussion posts differs.

In both cases, we observed that the general-purpose model used to calculate the similarity values of the discussions identified similar topical conversations. However, it did not identify the project issue specificity. Based on these findings, we observed that the {\tt RD-Detector} could fail to treat particular contexts of software projects. We will call this limitation the `project-specific limitation.'

\textit{Homebrew project}: The analysis of the first {\tt Homebrew}'s unrelated pair, \#3 (Table~\ref{tab:results_FP}), shows that the target discussion contains a link to the master discussion, `... for Homebrew mirror configurations. \#1906'. Link references can endorse or refute relationships between discussion posts. In this case, the text fragment that contains the link reference does not refute that the discussions are related; however, it also does not clearly emphasize that they are related. We analyzed the content of discussion pair \#3. We identified a limitation regarding the concept of `related discussions' that can directly influence the evaluation of the approach. We concluded that the interpretation of the `related discussions' concept depends on the evaluators' perspective. We call this limitation `concept imprecision.'

However, M\_Homebrew's feedback pointed out that pair \#3 is unrelated due to the `project-specific limitation.' The maintainer claims that both discussions address the same project feature, but they differ on the specificity of the issue addressed. According to M\_Homebrew maintainer, one discussion `...is asking what the policy is' and the other one `...is announcing support for a new feature. ' We also identified that the `project-specific limitation' justifies the other two false positives detected for the {\tt Homebrew} project. The maintainer justifies the unrelatedness of the pair \#4 (Table~\ref{tab:results_FP}) arguing that the strategy captured a high level of abstraction from the two discussions, `... problems with installing Homebrew.' However, the {\tt RD-Detector} did not capture the specificity of the problem, `...the problem that's encountered is different... .' Therefore, the relationship between the two is not confirmed because, according to M\_Homebrew `...each needs a different solution.'

We also analyzed the discussions' content of the third unrelated pair \#5 (Table~\ref{tab:results_FP}). Like the pair \#4, the discussions in \#5 address issues related to `Homebrew installation' but in different computing environments. Again, the {\tt RD-Detector} did not capture the specificity of the problem; however, both discussion posts address the same issue and have the project-related keywords.

\textit{Next.js project}: Analyzing the {\tt Next.js} false-positives, we note that the discussion creators (1) used screenshots to detail or describe the issues and (2) used error logs descriptions to show the stack trace of where the error took place. We also note the predominance of (3) template keywords and (4) project keywords in the false-positive discussion posts.

Users can add screenshots in the discussion body to help explain their problems. However, the {\tt RD-Detector} measures the semantic textual similarity between discussions content. The approach does not use images as a source of evidence. The discussion pairs \#7 and \#15 (Table~\ref{tab:results_FP}) exemplify this scenario. In addition, the master and target discussions use the same description template. After preprocessing, the templates' keywords may stand out against the real discussion content. We also identified the predominance of the template keywords in the false-positive pair \#16.

Discussions creators also use error logs or descriptions to describe the system's discrepancies or nonconformities. During the preprocessing phase, we remove error descriptions embedded in HTML tags. However, when users use error log content to express their questions, they usually intend to ask for help in solving a specific problem highlighted in the error log content. Removing the log also eliminates the problem specificity. The false-positive discussion pair \#11 (Table~\ref{tab:results_FP}) exemplifies the use of error log descriptions in discussion posts. However, project keywords in discussions of pair \#11 show similarities in the discussions' contents.

The analysis of the pairs \#13 and \#14 shows that (again) both discussions use the same description templates and have the same project keywords. Finally, the false-positive pairs \#6, \#8, \#9, \#10, \#12, \#17, \#18, \#19, \#20 (Table~\ref{tab:results_FP}) present the same set of keywords, which are limited due to the scope of the project. Keywords match lead to the `project-specific limitation.'

Based on maintainers' feedback, we can propose improvements to the proposed approach. For example, we can use the maintainers' judgments to optimize the classifier by providing samples of related and not related discussions. Furthermore, we can design strategies to minimize the project-specific limitation by treating the predominance of projects' keywords and templates' keywords.

\subsection{The implications of this research}

We envision this research enables opportunities for OSS communities and software engineering researchers as follows.

\textit{Long-term sustainability of \OSS communities}: the proposed approach increases the effectiveness of related discussions detection in GitHub Discussions. Their detection will no longer depend on the knowledge and availability of OSS maintainers. Maintainers can benefit from {\tt RD-Detector} to address the labor-intensive task of manually detecting related discussions and the time-intensive task of answering the same question multiple times. Such benefits enable maintainers to focus efforts on performing other activities to support the OSS community build and growth and the project sustainability, as pointed out by \cite{dias2021makes}. The precision values presented in Table~\ref{tab:result_precision} show that we can use a general-purpose deep learning model to detect related discussions in GitHub Discussions. Consequently, we aim to build an automated agent (Bot) to run the {\tt RD-Detector} algorithm and report occurrences of related discussion posts in the \gh repositories. In addition, we envision a practical application of our results in supporting the commenters' tasks on \gh. The commenters are contributors that enrich discussions in the OSS project by adding comments on collaborative conversation threads~\citep{canovas2022analysis}.

\textit{Knowledge sharing in \OSS communities}: due to the importance of knowledge sharing to OSS teams \citep{chen2013knowledge,tantisuwankul2019topological}, {\tt RD-Detector} can help maintainers reorganize project-related issues repeatedly discussed in OSS communities. According to the OSS maintainers' perspective,  the approach helps in identifying discussion threads that could be merged or made as comments on one another (Section~\ref{sec_results:mainteiners_perspective}), allowing users to find the correct answers in one place and avoid project knowledge decentralization \citep{ahasanuzzaman2016mining,silva2018duplicate}. In addition, managing related discussions can make finding the information community members are looking for easier since it enables reducing noise occurrences and helps valuable discussions posts become `more visible' \citep{mamykina2011design}.

\textit{\OSS communities' growth}: the remarkable acceptance of \GHD in OSS projects \citep{Evi21_mainteiners} creates expectations about the forum growth. As \GHD grows and becomes popular, it faces the same limitations and problems that other similar forums, such as related discussions (duplicate or near-duplicate) occurrence~\citep{zhang2015multi_DupPredictor,ahasanuzzaman2016mining,zhang2017detecting,silva2018duplicate,wang2020duplicate,pei2021attention}. Identifying related threads can help users find previous posts already asked and answered, reducing the waste of time while waiting for feedback~\citep{ahasanuzzaman2016mining} and the fear of submitting duplicates \citep{ford2016paradise}. It also helps to suggest related posts before creating a new thread~\citep{zhang2017detecting}, assisting users to find posts with similar issues~\citep{ahasanuzzaman2016mining,zhang2017detecting}, helping new users asking better questions \citep{ford2018somentoring}, and reducing the manual responsibility of maintainers to sustain a healthy community~\citep{guizani2022attracting}. Besides, every online community depends on volunteer activity, so it is essential to attract and retain new members to guarantee community sustainability~\citep{cho2021potential,guizani2022attracting}. However, it is well-known that newcomers face barriers to onboard to OSS communities~\citep{steinmacher2015social}. Given that the \GHD is a rich project source of information, newcomers and inexperienced users also may benefit from identifying related posts in the forum. Such users can analyze the related discussion threads to get different formulations of the same problem and better understand a project-related issue \citep{abric2019can}. Through a cognitive process called `social or collaborative reflection,' newcomers and inexperienced users can learn from the reflection that others registered in the related discussion threads~\citep{prilla2020does}. In this way, our approach can support the long-term growth and maintenance of the \GHD forum.

\textit{OSS Maintainers}: previous research report that OSS maintainers need to manage multiple aspects of the project to ensure that the project vision endures \citep{guizani2022attracting} and the projects' long-term sustainability \citep{dias2021makes}. To do so, maintainers' perform different activities encompassing code and non-coding tasks \citep{dias2021makes,trinkenreich2021pots}. Such responsibilities intensify the workload of OSS maintainers. \cite{tan2020scaling} highlight proposing tools as a best practice to decentralize maintainers' responsibilities. In addition, \cite{dias2021makes} report that `to sustain a long term vision of the project, maintainers should delegate tasks.' In this way, the {\tt RD-Detector} emerges as a tool to alleviate the maintainers' labor-intensive task of detecting related discussion occurrences and managing duplicates or near-duplicate discussions.

\textit{Research Opportunities}: to address the diversity and uniqueness of OSS communities hosted on \gh, the {\tt RD-Detector} is based on a Sentence-BERT general-purpose machine learning model. Besides, the approach uses descriptive statistics to calculate the local threshold used to detect related discussion pairs. Given that software projects are unique development ecosystems, we can not use a single and universal number to detect related discussion pairs. Our results (Table~\ref{tab:result_precision}) endorse the effectiveness of the {\tt RD-Detector}. However, we do not know in advance how many or which discussions are genuinely related in the \GHD forum, which prevents us from measuring the approach recall rate. This gap brings opportunities for future work.

\section{Limitations}
\label{sec_limitations}

Although we proposed a parameterizable approach based on general-purpose machine learning models and descriptive statistics, this research may likely present limitations.

The research's main limitation relates to the generalization of the results. We are aware of the diversity and uniqueness of the OSS communities hosted on GitHub. Since we assessed {\tt RD-Detector} over three OSS communities, we cannot guarantee whether our findings generalize to all projects. GitHub's professionals (coauthors in this research) singled out the selected OSS communities to minimize this limitation. In addition, the dataset refers to a specific time window that does not reflect the current moment of GitHub Discussions. However, the {\tt Next.js} project stands out as it has a high rate of use of the forum. More experiments are needed to assess the results in different OSS communities.

Moreover, judging the relatedness between discussions is a subjective task and could introduce biases in reporting the {\tt RD-Detector} effectiveness. The {\tt RD-Detector} evaluation presents some challenges: (1) people who judge the relatedness of the discussions need to semantically analyze the content of the posts; (2) judging the technical aspects of the discussion posts requires prior knowledge about the project; and (3) the judgment involves human (in)precision regarding the concept of relatedness, although we defined the concept of related and duplicate discussions, the interpretation depends on the evaluators' perspective. To minimize this threat, we introduced the `related discussions' meaning to OSS maintainers and SE researchers before classifying related discussion candidates. We also contacted selected OSS maintainers to judge the related discussion candidates.

The local threshold calculation can also be a limitation. We consider related discussion candidates those pairs identified as outliers in a distribution $S$. $S$ contains the similarity values of the $K$ most similar target discussions for each discussion post in the dataset. As we increase the value of $K$, the median of the distribution decreases, and so does the local threshold value. However, we focused on improving the {\tt RD-Detector} precision. Higher precision values ensure greater assertiveness in detecting related discussions. So, we set the $K$ value to 5 and 10.

Finally, it was not possible to measure the {\tt RD-Detector}'s recall rate. To the best of our knowledge, this is the first work detecting related discussions in the \GHD forum. However, achieved precision values ( Table~\ref{tab:result_precision} ) show the effectiveness of the {\tt RD-Detector} in detecting related discussions. Maintainers benefit from the {\tt RD-Detector} quality by decreasing the time spent reviewing related discussion candidates. 

\section{Conclusion}
\label{sec_conclusion}

In this work, we presented the {\tt RD-Detector}, an approach to detect related discussion candidates in the \GHD forum. We assessed {\tt RD-Detector} over public discussions collected from three OSS communities. In total, the approach evaluated the semantic similarity of 11,162 discussion posts. OSS maintainers and SE researchers judged the detected related discussion candidates. They classified pairs of related discussion candidates as duplicates, related or not related. We measured the {\tt RD-Detector} precision rate using OSS maintainers' and SE researchers' judgment. Our results show that we can use a general-purpose deep machine learning model applicable to NLP problems to detect related discussions in the \GHD forum. The {\tt RD-Detector} achieved an average precision rate of 90.11\%, considering the top-10 most similar discussion pairs for each discussion post in the dataset $D$ ($K=10$).

The {\tt RD-Detector} uses a Sentence-BERT (SBERT) pre-trained general-purpose model to compute semantically significant sentence embeddings of discussion posts and detect related discussion candidates. Using publicly available machine learning models brings flexibility to the approach. As researchers release new exchangeable models, one can update the {\tt RD-Detector}. Besides, the {\tt RD-Detector} calculates local threshold values to detect related discussion candidates in different OSS communities. We use descriptive statistics to calculate the upper inner fence value of a distribution containing the similarity values data of the $K$ most similar target discussions to each discussion under processing. We consider related discussions those identified as outliers in the distribution. 

The approach's outputs were unique (Tables~\ref{tab:descripStat_Gatsby}, \ref{tab:decripStat_Homebrew}, and \ref{tab:descripStat_NextJS}). Our results highlight the need for strategies designed to quickly adapt to the dynamism with which OSS communities grow and change. We found that OSS communities create related discussions to emphasize their need for help, add new information on previous discussions threads, and ask for alternative solutions. From the maintainers' perspective, we also found that related discussions have the same resolution and similar problems, address additional issues, and address different specifics within the same topic. 

Maintainers can benefit from {\tt RD-Detector} to address the labor-intensive task of manually detecting related discussions and the time-intensive task of answering the same question multiple times. In addition, OSS maintainers can benefit from the approach to prioritize the development or update project-related issues frequently discussed, understand why users duplicate questions, control the propagation of duplicates, and support the project knowledge sharing. According to the maintainers' feedback, one can merge related discussion threads or make related discussions as comments on one another.

We reported and discussed our results with the professional \gh team (co-authors of this work). Our findings showed a real need to plan and tackle related discussions on GitHub Discussions. Consequently, the \gh engineering team is testing some changes to the Discussions interface. In addition to providing the discussion title, body text, and category, users would have to confirm they have searched for similar threads before creating new posts (using a checkbox).

As the next step, we intend to implement {\tt RD-Detector} as a bot to run over the \GHD data and fetch, analyze, detect, and report related discussions occurrences. We believe that this research also brings opportunities to enable project knowledge acquisition and transfer by providing users with project-related issues and by making the projects' knowledge easy to find. In addition, our results can enable project knowledge reuse as users can access related discussions that have already been asked and answered and the project knowledge categorization by identifying similar documents.

\begin{acknowledgements}

We would like to thank GitHub and the OSS maintainers for supporting this research. We also would like to thank the financial support granted by CNPq through processes number 314174/ 2020-6 and 313067/2020-1. CAPES financial code 001. FAPESP under grant 2020/05191-2. FAPEAM through process number 062.00150/2020. This research was also carried out within the scope of the Samsung-UFAM Project for Education and Research (SUPER), according to Article 48 of Decree number 6.008/2006(SUFRAMA).

\end{acknowledgements}

\section*{Declaration of competing interest}

The fourth and the fifth authors of this manuscript participate in the GitHub Discussions engineering team. The other authors have no competing interests to declare that are relevant to the content of this article.

%
%

\bibliographystyle{apalike}
\bibliography{01_main}

\begin{thebibliography}{}

\bibitem[Abric et~al., 2019]{abric2019can}
Abric, D., Clark, O.~E., Caminiti, M., Gallaba, K., and McIntosh, S. (2019).
\newblock Can duplicate questions on stack overflow benefit the software
  development community?
\newblock In {\em 2019 IEEE/ACM 16th International Conference on Mining
  Software Repositories (MSR)}, pages 230--234. IEEE.

\bibitem[Agirre et~al., 2015]{agirre-etal-2015-semeval}
Agirre, E., Banea, C., Cardie, C., Cer, D., Diab, M., Gonzalez-Agirre, A., Guo,
  W., Lopez-Gazpio, I., Maritxalar, M., Mihalcea, R., Rigau, G., Uria, L., and
  Wiebe, J. (2015).
\newblock {S}em{E}val-2015 task 2: Semantic textual similarity, {E}nglish,
  {S}panish and pilot on interpretability.
\newblock In {\em Proceedings of the 9th International Workshop on Semantic
  Evaluation ({S}em{E}val 2015)}, pages 252--263, Denver, Colorado. Association
  for Computational Linguistics.

\bibitem[Ahasanuzzaman et~al., 2016]{ahasanuzzaman2016mining}
Ahasanuzzaman, M., Asaduzzaman, M., Roy, C.~K., and Schneider, K.~A. (2016).
\newblock Mining duplicate questions of stack overflow.
\newblock In {\em 2016 IEEE/ACM 13th Working Conference on Mining Software
  Repositories (MSR)}, pages 402--412. IEEE.

\bibitem[Calefato et~al., 2021]{calefato2021will}
Calefato, F., Gerosa, M.~A., Iaffaldano, G., Lanubile, F., and Steinmacher, I.
  (2021).
\newblock Will you come back to contribute? investigating the inactivity of oss
  core developers in github.
\newblock {\em arXiv preprint arXiv:2103.04656}.

\bibitem[C{\'a}novas~Izquierdo and Cabot, 2022]{canovas2022analysis}
C{\'a}novas~Izquierdo, J.~L. and Cabot, J. (2022).
\newblock On the analysis of non-coding roles in open source development.
\newblock {\em Empirical Software Engineering}, 27(1):1--32.

\bibitem[Chen et~al., 2013]{chen2013knowledge}
Chen, X., Li, X., Clark, J.~G., and Dietrich, G.~B. (2013).
\newblock Knowledge sharing in open source software project teams: A
  transactive memory system perspective.
\newblock {\em International Journal of Information Management},
  33(3):553--563.

\bibitem[Cho and Wash, 2021]{cho2021potential}
Cho, J. and Wash, R. (2021).
\newblock How potential new members approach an online community.
\newblock {\em Computer Supported Cooperative Work (CSCW)}, 30(1):35--77.

\bibitem[Cohen, 1960]{cohen1960coefficient}
Cohen, J. (1960).
\newblock A coefficient of agreement for nominal scales.
\newblock {\em Educational and psychological measurement}, 20(1):37--46.

\bibitem[Community, 2022]{gatsby_doc}
Community, G. (2022).
\newblock Gatsby v4. url: https://github.com/gatsbyjs/gatsby/
  blob/master/readme.md .accessed 23 january 2022.

\bibitem[Dias et~al., 2021]{dias2021makes}
Dias, E., Meirelles, P., Castor, F., Steinmacher, I., Wiese, I., and Pinto, G.
  (2021).
\newblock What makes a great maintainer of open source projects?
\newblock In {\em 2021 IEEE/ACM 43rd International Conference on Software
  Engineering (ICSE)}, pages 982--994. IEEE.

\bibitem[Face, 2021]{hugginface}
Face, H. (2021).
\newblock sentence-transformers/all-mpnet-base-v2.
  url:https://huggingface.co/sentence-transformers/all-mpnet-base-v2 . accessed
  13 january 2022.

\bibitem[Ford et~al., 2018]{ford2018somentoring}
Ford, D., Lustig, K., Banks, J., and Parnin, C. (2018).
\newblock "we don't do that here": How collaborative editing with mentors
  improves engagement in social q\&a communities.
\newblock In {\em Proceedings of the 2018 CHI Conference on Human Factors in
  Computing Systems}, CHI '18, page 1–12, New York, NY, USA. Association for
  Computing Machinery.

\bibitem[Ford et~al., 2016]{ford2016paradise}
Ford, D., Smith, J., Guo, P.~J., and Parnin, C. (2016).
\newblock Paradise unplugged: Identifying barriers for female participation on
  stack overflow.
\newblock In {\em Proceedings of the 2016 24th ACM SIGSOFT International
  Symposium on Foundations of Software Engineering}, FSE 2016, page 846–857,
  New York, NY, USA. Association for Computing Machinery.

\bibitem[GitHub, 2021a]{discussion}
GitHub, I. (2021a).
\newblock Github discussions. url: https://docs.github.com/en/ discussions.
  accessed 23 january 2022.

\bibitem[GitHub, 2021b]{discussion_category}
GitHub, I. (2021b).
\newblock Managing categories for discussions in your repository. url:
  https://docs.github.com/en/discussions/managing-discussions-for-your-community/managing-categories-for-discussions-in-your-repository.
  accessed 23 january 2022.

\bibitem[GitHub, 2021c]{discussion_search}
GitHub, I. (2021c).
\newblock Searching discussions. url:
  https://docs.github.com/en/github/searching-for-information-on-github/searching-discussions.
  accessed 23 january 2022.

\bibitem[GitHub, 2022]{resources_GH}
GitHub, I. (2022).
\newblock What is github discussions? a complete guide. url:
  https://resources.github.com/devops/ process/planning/discussions/. accessed
  21 april 2022.

\bibitem[Guizani et~al., 2022]{guizani2022attracting}
Guizani, M., Zimmermann, T., Sarma, A., and Ford, D. (2022).
\newblock Attracting and retaining oss contributors with a maintainer
  dashboard.
\newblock {\em arXiv preprint arXiv:2202.07740}.

\bibitem[Hata et~al., 2022]{hata2022github}
Hata, H., Novielli, N., Baltes, S., Kula, R.~G., and Treude, C. (2022).
\newblock Github discussions: An exploratory study of early adoption.
\newblock {\em Empirical Software Engineering}, 27(1):1--32.

\bibitem[Kim et~al., 2005]{kim2005improving}
Kim, Y., Lee, S., Dollmann, M., and Geierhos, M. (2005).
\newblock Improving classifiers for semantic annotation of software
  requirements with elaborate syntatic structure.
\newblock {\em International Journal of Advanced Science and Technology, ISSN},
  4238:123--136.

\bibitem[Landis and Koch, 1977]{landis1977measurement}
Landis, J.~R. and Koch, G.~G. (1977).
\newblock The measurement of observer agreement for categorical data.
\newblock {\em biometrics}, pages 159--174.

\bibitem[Lee and Shin, 2020]{lee2020machine}
Lee, I. and Shin, Y.~J. (2020).
\newblock Machine learning for enterprises: Applications, algorithm selection,
  and challenges.
\newblock {\em Business Horizons}, 63(2):157--170.

\bibitem[Li et~al., 2018]{li2018issue}
Li, L., Ren, Z., Li, X., Zou, W., and Jiang, H. (2018).
\newblock How are issue units linked? empirical study on the linking behavior
  in github.
\newblock In {\em 2018 25th Asia-Pacific Software Engineering Conference
  (APSEC)}, pages 386--395. IEEE.

\bibitem[Li et~al., 2017]{li2017detecting}
Li, Z., Yin, G., Yu, Y., Wang, T., and Wang, H. (2017).
\newblock Detecting duplicate pull-requests in github.
\newblock In {\em Proceedings of the 9th Asia-Pacific Symposium on
  Internetware}, pages 1--6.

\bibitem[Li et~al., 2020]{li2020redundancy}
Li, Z., Yu, Y., Zhou, M., Wang, T., Yin, G., Lan, L., and Wang, H. (2020).
\newblock Redundancy, context, and preference: An empirical study of duplicate
  pull requests in oss projects.
\newblock {\em IEEE Transactions on Software Engineering}.

\bibitem[Liu, 2021]{Evi21_mainteiners}
Liu, E. (2021).
\newblock Github discussions is out of beta. url:
  https://github.blog/2021-08-17-github-discussions-out-of-beta/. accessed 15
  january 2022.

\bibitem[Liu, 2022]{Evi20}
Liu, E. (2022).
\newblock How five open source communities are using github discussions.
  url:https://github.blog/2022-01-13-how-five-open-source-
  communities-are-using-github-discussions/.

\bibitem[Liu, 2011]{liu2011learning}
Liu, T.-Y. (2011).
\newblock Learning to rank for information retrieval.

\bibitem[Mamykina et~al., 2011]{mamykina2011design}
Mamykina, L., Manoim, B., Mittal, M., Hripcsak, G., and Hartmann, B. (2011).
\newblock Design lessons from the fastest q\&a site in the west.
\newblock In {\em Proceedings of the SIGCHI Conference on Human Factors in
  Computing Systems}, CHI '11, page 2857–2866, New York, NY, USA. Association
  for Computing Machinery.

\bibitem[Niyogi, 2020]{Niyogi20}
Niyogi, S. (2020).
\newblock New from satellite 2020: Github discussions, codespaces, securing
  code in private repositories, and more. url:
  https://github.blog/2020-05-06-new-from-satellite-2020-github-codespaces-github-
  discussions-securing-code-in-private-repositories-and-more/. accessed 21
  january 2022.

\bibitem[Pei et~al., 2021]{pei2021attention}
Pei, J., Wu, Y., Qin, Z., Cong, Y., and Guan, J. (2021).
\newblock Attention-based model for predicting question relatedness on stack
  overflow.
\newblock In {\em 2021 IEEE/ACM 18th International Conference on Mining
  Software Repositories (MSR)}, pages 97--107. IEEE.

\bibitem[P{\'e}rez-Soler et~al., 2018]{perez2018collaborative}
P{\'e}rez-Soler, S., Guerra, E., and de~Lara, J. (2018).
\newblock Collaborative modeling and group decision making using chatbots in
  social networks.
\newblock {\em IEEE Software}, 35(6):48--54.

\bibitem[Polyzotis et~al., 2017]{polyzotis2017data}
Polyzotis, N., Roy, S., Whang, S.~E., and Zinkevich, M. (2017).
\newblock Data management challenges in production machine learning.
\newblock In {\em Proceedings of the 2017 ACM International Conference on
  Management of Data}, pages 1723--1726.

\bibitem[Prilla et~al., 2020]{prilla2020does}
Prilla, M., Blunk, O., and Chounta, I.-A. (2020).
\newblock How does collaborative reflection unfold in online communities? an
  analysis of two data sets.
\newblock {\em Computer Supported Cooperative Work (CSCW)}, 29(6):697--741.

\bibitem[Project, 2022]{homebrew_doc}
Project, H. (2022).
\newblock Homebrew documentation. url:https://docs.brew.sh/. accessed 23
  january 2022.

\bibitem[Reimers, 2021]{SBERT_site}
Reimers, N. (2021).
\newblock Sentencetransformers documentation. url: https://www.sbert.net/.
  accessed 15 january 2022.

\bibitem[Reimers and Gurevych, 2019]{reimers2019sentence}
Reimers, N. and Gurevych, I. (2019).
\newblock Sentence-bert: Sentence embeddings using siamese bert-networks.
\newblock {\em arXiv preprint arXiv:1908.10084}.

\bibitem[Ren et~al., 2019]{ren2019identifying}
Ren, L., Zhou, S., Kastner, C., and Wasowski, A. (2019).
\newblock Identifying redundancies in fork-based development.
\newblock In {\em 2019 IEEE 26th International Conference on Software Analysis,
  Evolution and Reengineering (SANER)}, pages 230--241. IEEE.

\bibitem[Schelter et~al., 2018]{schelter2018challenges}
Schelter, S., Biessmann, F., Januschowski, T., Salinas, D., Seufert, S., and
  Szarvas, G. (2018).
\newblock On challenges in machine learning model management.

\bibitem[Silva et~al., 2018]{silva2018duplicate}
Silva, R.~F., Paix{\~a}o, K., and de~Almeida~Maia, M. (2018).
\newblock Duplicate question detection in stack overflow: A reproducibility
  study.
\newblock In {\em 2018 IEEE 25th international conference on software analysis,
  evolution and reengineering (SANER)}, pages 572--581. IEEE.

\bibitem[Sirres et~al., 2018]{sirres2018augmenting}
Sirres, R., Bissyand{\'e}, T.~F., Kim, D., Lo, D., Klein, J., Kim, K., and
  Traon, Y.~L. (2018).
\newblock Augmenting and structuring user queries to support efficient
  free-form code search.
\newblock {\em Empirical Software Engineering}, 23(5):2622--2654.

\bibitem[Steinmacher et~al., 2015]{steinmacher2015social}
Steinmacher, I., Conte, T., Gerosa, M.~A., and Redmiles, D. (2015).
\newblock Social barriers faced by newcomers placing their first contribution
  in open source software projects.
\newblock In {\em Proceedings of the 18th ACM conference on Computer supported
  cooperative work \& social computing}, pages 1379--1392.

\bibitem[Storey et~al., 2014]{storey2014r}
Storey, M.-A., Singer, L., Cleary, B., Figueira~Filho, F., and Zagalsky, A.
  (2014).
\newblock The (r) evolution of social media in software engineering.
\newblock In {\em Future of Software Engineering Proceedings}, pages 100--116.

\bibitem[Storey et~al., 2016]{storey2016social}
Storey, M.-A., Zagalsky, A., Figueira~Filho, F., Singer, L., and German, D.~M.
  (2016).
\newblock How social and communication channels shape and challenge a
  participatory culture in software development.
\newblock {\em IEEE Transactions on Software Engineering}, 43(2):185--204.

\bibitem[Tan and Zhou, 2020]{tan2020scaling}
Tan, X. and Zhou, M. (2020).
\newblock Scaling open source software communities: Challenges and practices of
  decentralization.
\newblock {\em IEEE Software}, 39(1):70--75.

\bibitem[Tantisuwankul et~al., 2019]{tantisuwankul2019topological}
Tantisuwankul, J., Nugroho, Y.~S., Kula, R.~G., Hata, H., Rungsawang, A.,
  Leelaprute, P., and Matsumoto, K. (2019).
\newblock A topological analysis of communication channels for knowledge
  sharing in contemporary github projects.
\newblock {\em Journal of Systems and Software}, 158:110416.

\bibitem[Trinkenreich et~al., 2021]{trinkenreich2021pots}
Trinkenreich, B., Guizani, M., Wiese, I.~S., Conte, T., Gerosa, M., Sarma, A.,
  and Steinmacher, I. (2021).
\newblock Pots of gold at the end of the rainbow: What is success for open
  source contributors.
\newblock {\em IEEE Transactions on Software Engineering}.

\bibitem[Tukey et~al., 1977]{tukey1977exploratory}
Tukey, J.~W. et~al. (1977).
\newblock {\em Exploratory data analysis}, volume~2.
\newblock Reading, Mass.

\bibitem[Vercel, 2022]{nextJS_doc}
Vercel, I. (2022).
\newblock Create a next.js app.
  url:https://nextjs.org/learn/basics/create-nextjs-app.accessed 23 january
  2022.

\bibitem[Wang et~al., 2020]{wang2020duplicate}
Wang, L., Zhang, L., and Jiang, J. (2020).
\newblock Duplicate question detection with deep learning in stack overflow.
\newblock {\em IEEE Access}, 8:25964--25975.

\bibitem[Yazdaninia et~al., 2021]{yazdaninia2021characterization}
Yazdaninia, M., Lo, D., and Sami, A. (2021).
\newblock Characterization and prediction of questions without accepted answers
  on stack overflow.
\newblock In {\em 2021 IEEE/ACM 29th International Conference on Program
  Comprehension (ICPC)}, pages 59--70. IEEE.

\bibitem[Yu et~al., 2018]{yu2018dataset}
Yu, Y., Li, Z., Yin, G., Wang, T., and Wang, H. (2018).
\newblock A dataset of duplicate pull-requests in github.
\newblock In {\em Proceedings of the 15th International Conference on Mining
  Software Repositories}, pages 22--25.

\bibitem[Zhang et~al., 2017]{zhang2017detecting}
Zhang, W.~E., Sheng, Q.~Z., Lau, J.~H., and Abebe, E. (2017).
\newblock Detecting duplicate posts in programming qa communities via latent
  semantics and association rules.
\newblock In {\em Proceedings of the 26th International Conference on World
  Wide Web}, pages 1221--1229.

\bibitem[Zhang et~al., 2015]{zhang2015multi_DupPredictor}
Zhang, Y., Lo, D., Xia, X., and Sun, J.-L. (2015).
\newblock Multi-factor duplicate question detection in stack overflow.
\newblock {\em Journal of Computer Science and Technology}, 30(5):981--997.

\bibitem[Zhang et~al., 2020]{zhang2020ilinker}
Zhang, Y., Wu, Y., Wang, T., and Wang, H. (2020).
\newblock ilinker: a novel approach for issue knowledge acquisition in github
  projects.
\newblock {\em World Wide Web}, 23(3):1589--1619.

\bibitem[Zhou et~al., 2017]{zhou2017machine}
Zhou, L., Pan, S., Wang, J., and Vasilakos, A.~V. (2017).
\newblock Machine learning on big data: Opportunities and challenges.
\newblock {\em Neurocomputing}, 237:350--361.

\end{thebibliography}

\end{document}